\documentclass[twoside,11pt]{article}

\usepackage{graphicx, subfigure}
\usepackage[top=1in,bottom=1in,left=1in,right=1in]{geometry}
\usepackage[sort&compress]{natbib} 
\usepackage{amsfonts, amsmath, amssymb, amsthm, constants, bbm}
\usepackage{MnSymbol}
\usepackage{authblk}
\usepackage{hyperref}
\usepackage{enumitem}

\usepackage{color}
\definecolor{darkred}{RGB}{100,0,0}
\definecolor{darkgreen}{RGB}{0,100,0}
\definecolor{darkblue}{RGB}{0,0,150}
\definecolor{orange}{RGB}{200,100,0}

\usepackage{hyperref}
\hypersetup{colorlinks=true, linkcolor=darkred, citecolor=darkgreen, urlcolor=darkblue}
\usepackage{url}

\newtheorem{thm}{Theorem}
\newtheorem{prp}{Proposition}
\newtheorem{lem}{Lemma}

\theoremstyle{remark}
\newtheorem{rem}{Remark}
\newtheorem{ex}{Example}

\def\beq{\begin{equation}} % \setcounter{equation}{1}}
\def\eeq{\end{equation}}
\def\beqn{\begin{eqnarray*}}
\def\eeqn{\end{eqnarray*}}
\def\Bitem{\begin{itemize}\setlength{\itemsep}{.2in}}
\def\bitem{\begin{itemize}\setlength{\itemsep}{.05in}}
\def\eitem{\end{itemize}}
\def\Benum{\begin{enumerate}\setlength{\itemsep}{.2in}}
\def\benum{\begin{enumerate}\setlength{\itemsep}{.05in}}
\def\eenum{\end{enumerate}}
\def\bmult{\begin{multline*}}
\def\emult{\end{multline*}}
\def\bcenter{\begin{center}}
\def\ecenter{\end{center}}
\def\bframe{\begin{frame}}
\def\eframe{\end{frame}}

\newcommand{\thmref}[1]{Theorem~\ref{thm:#1}}
\newcommand{\prpref}[1]{Proposition~\ref{prp:#1}}

\newcommand{\lemref}[1]{Lemma~\ref{lem:#1}}
\newcommand{\secref}[1]{Section~\ref{sec:#1}}
\newcommand{\figref}[1]{Figure~\ref{fig:#1}}

\def\cC{\mathcal{C}}

\def\cN{\mathcal{N}}

\def\cS{\mathcal{S}}

\def\cV{\mathcal{V}}

\def\bR{\mathbf{R}}

\def\bX{\mathbf{X}}

\def\br{\mathbf{r}}

\def\bx{\mathbf{x}}

\def\bbD{\mathbb{D}}

\def\bbN{\mathbb{N}}

\def\bbR{\mathbb{R}}
\def\bbS{\mathbb{S}}

\newcommand{\E}{\operatorname{\mathbb{E}}}
\renewcommand{\P}{\operatorname{\mathbb{P}}}
\newcommand{\Var}{\operatorname{Var}}
\newcommand{\Cov}{\operatorname{Cov}}

\def\eps{\varepsilon}

\DeclareMathOperator{\sign}{sign}

\newcommand{\IND}[1]{\mathbbm{1}_{\{ #1 \}}}

\def\scan{{\text{\sc scan}}}

\definecolor{purple}{rgb}{0.4,.1,.9}
\begin{document}
\thispagestyle{empty}

\title{Distribution-Free Detection of Structured Anomalies: \\ Permutation and Rank-Based Scans}
\author[1]{Ery Arias-Castro}
\author[2]{Rui M. Castro}
\author[2]{Ervin T\'anczos}
\author[1]{Meng Wang}
\affil[1]{University of California, San Diego}
\affil[2]{Technische Universiteit Eindhoven} 
\date{}
\maketitle

\begin{abstract}
The scan statistic is by far the most popular method for anomaly detection, being popular in syndromic surveillance, signal and image processing, and target detection based on sensor networks, among other applications.  The use of the scan statistics in such settings yields a hypothesis testing procedure, where the null hypothesis corresponds to the absence of anomalous behavior.  If the null distribution is known, then calibration of a scan-based test is relatively easy, as it can be done by Monte Carlo simulation.  When the null distribution is unknown, it is less straightforward.  
We investigate two procedures.  The first one is a calibration by permutation and the other is a rank-based scan test, which is distribution-free and less sensitive to outliers.  Furthermore, the rank scan test requires only a one-time calibration for a given data size making it computationally much more appealing.  In both cases, we quantify the performance loss with respect to an oracle scan test that knows the null distribution.  We  show that using one of these calibration procedures results in only a very small loss of power in the context of a natural exponential family. This includes the classical normal location model, popular in signal processing, and the Poisson model, popular in syndromic surveillance.  We perform numerical experiments on simulated data further supporting our theory and also on a real dataset from genomics. 
\end{abstract}

\section{Introduction} \label{sec:intro}

Signal detection (and localization) is important in a large variety of applications, encompassing any situation where the goal is to discover patterns or detect/locate anomalies.  Our focus is on the detection of anomalous behavior which is endowed with some structure.  For instance, one might have data consisting of the physical location of a sensor and the corresponding measurement, and would like to determine if there is a spatial region where measurements are unusually high \citep{balakrishnan:02}.   A standard way to tackle this problem is the use of a scan statistic which essentially inspects all (or at least a large number of) possible anomalous patterns.  It usually corresponds to a form of generalized likelihood ratio test  \citep{Kul}.  In \citep{cheung:13} the scan statistic was used to detect small geographic areas with large suicide rates and \citep{guerriero:09} used the scan statistic for target detection using distributed sensors in a two dimensional region.
Although computationally this approach might be challenging, there are a number of situations where it is possible to compute the scan statistic in nearly linear time \citep{neill2004rapid,neill2012fast,MGD,walther2010optimal}.
 
For the purpose of illustration, consider the following prototypical example\footnote{In fact, this setting might have been the original motivation for the work on the scan statistic \citep{wallenstein2009joseph}.}:  suppose we have event data over a certain time period and want to detect if there is a time interval with an unusually high concentration of events.  To make things more concrete and move towards the setting we consider in this paper, assume one can model these event data as a realization of a Poisson process and bin the data, so that we observe a sequence of Poisson random variables. The scan statistic in this particular case combines sums of these values over (discrete) intervals of different sizes and location, together with some normalization --- see \eqref{scan} further down.  In this scenario we want to perform a hypotheses test, where the null hypothesis is that no anomaly is present (a homogenous Poisson process) versus the alternative where some intervals have an elevated rate of events (an inhomogenous process).  If the (constant) rate is known under the null, then the null distribution is completely specified and the test can be calibrated either analytically or by Monte Carlo simulation.  But what if the null event rate is unknown?  What are possible ways to properly calibrate the test?  What is the price to pay in terms of power?

One can regard the scan statistic as a comparison between observations in one interval to those outside the interval.  This point of view leads naturally to a two-sample problem for each interval, which is then followed by some form of multiple testing since we scan many intervals.  Thus drawing from the classical literature on the two-sample problem, two approaches can be considered:
\bitem
\item {\em Calibration by permutation.}  This amounts to using the permutation distribution of the scan statistic for inference (detection/estimation).   
\item {\em Scanning the ranks.}  This amounts to replacing each observation with its rank before scanning.  Calibration of such a test can be done by Monte Carlo simulation before the observation of data, as long as the size of the data is known.
\eitem

The perspective offered by the two-sample testing framework makes these two procedures very natural.  
The permutation scan has been suggested in a number of papers and applied in a number of ways in different contexts.  
It is a standard approach in neuroimaging \citep{nichols2002nonparametric} and is suggested in syndromic surveillance \citep{kulldorff2005stp,19843331,huang2007spatial}.  It was suggested by \cite{walther2010optimal} in the context of a sensor network with binary output and by \cite{MR2792412} in the context of detecting a change in a sequence of images. 

Surprisingly enough, the method based on ranks appears to be relatively new in the present context.  It was specifically (and simultaneously) proposed as a standalone procedure by \cite{jung2015nonparametric}\footnote{This article was made public after our paper was posted on the {\tt arxig.org}.  To the best of our knowledge, this other publication became publicly available on October 20, 2015 ({\tt doi:10.1186/s12942-015-0024-6}), a couple of months after ours appeared online on August 12, 2015.}, where the authors compute the scan statistic on ranks instead of the data itself.  Nevertheless, rank-based methodologies have been used earlier in similar settings, but with a different purpose in mind.  For instance, the use of ranks in the context of the scan statistic also appears in \citep{mcfowland2013fast} through the computation of empirical P-values.  It is important to note that the use of ranks in the last reference is of a rather different nature than that we propose in our work, and that the emphasis in that paper is on the ability to efficiently compute/approximate scan statistics, while in our work the emphasis is on the calibration of scan tests when the null distribution is not known.

Although less popular, as in the two-sample testing setting, a procedure based on ranks offers some significant advantages over calibration by permutation: (i) it is more robust to outliers and ; (ii) its calibration can be done by Monte Carlo simulation and requires only the knowledge of the sample size\footnote{The latter explains why, in two-sample testing, methods based on ranks were feasible decades before methods based on permutations, which typically require access to a computer.}.  Point (ii) is rather pertinent, as computationally this is a huge advantage over calibration by permutation. Furthermore, this property is rather advantageous if one desires to apply the test repeatedly on several datasets of same size; compare with a calibration by permutation: typically, several hundred permutations are sampled at random and, for each one of them, the scan statistic is computed, and all this is done each time the test is applied.

In this paper we study the performance of both the permutation and rank scan methods, providing strong asymptotic guarantees as well as insights on the their finite-sample performance in some numerical experiments.
In the context of a natural exponential family --- which includes the classical normal location model and the Poisson example above --- we find that the permutation scan test and the rank scan test come very close to performing as well as the oracle scan test, which we define as the scan test calibrated by Monte Carlo with (clairvoyant) knowledge of the null distribution. 
We perform numerical experiments on simulated data which confirm our theory, and also some experiments using a real dataset from genomics.  

As specified below, we focus on a ``static'' setting, where the length of the signal being monitored is fixed a priori.  Adding time is typically done by adding one `dimension' to the framework, as done for example in \citep{kulldorff2005stp}.

\subsection{General setting}

A typical framework for static anomaly detection --- which includes detection in digital signals and images, sensor networks, biological data, and more --- may be described in general terms as follows.   We observe a set of independent random variables, denoted $(X_v : v \in \cV)$, where $\cV$ is a finite index set of size $N$.  This is a snapshot of the state of the environment, where each element of $\cV$ corresponds to an element of the environment (e.g., these correspond to nodes of a network, pixels in an image, genes, etc.).
In this work we take a hypothesis testing point of view. Under the null hypothesis, corresponding to the nominal state when no anomalies are present, these random variables are Independent and Identically Distributed (IID) with distribution $F_0$.  Under the alternative, some of these random variables will have a different distribution.  Formally, let $\bbS \subset 2^\cV$ denote a class of possibly anomalous subsets, corresponding to the anomalous patterns we expect to encounter (this would be a class of intervals in the example that we used earlier).  Under the alternative hypothesis there is a subset $\cS \in \bbS$ such that, for each $v \in \cS$, $X_v \sim F_v$ for some distributions $F_v \ne F_0$, and independent of $(X_v : v \in \cV \setminus \cS)$, which are still IID with distribution $F_0$.
In a number of important applications the variables are real-valued and the anomalous variables take larger-than-usual values, which can be formalized by the assumption that each $F_v$ stochastically dominates\footnote{For two distribution (functions) on the real line, $F$ and $G$, we say that $G$ stochastically dominates $F$ if $G(t) \le F(t)$ for all $t \in \bbR$.  We denote this by $G \succeq F$.} $F_0$.  We take this to be the case throughout most of the paper.  While the standard scan test is calibrated by Monte Carlo by repeated sampling from the null distribution $F_0$, in contrast, the procedures we study here --- the permutation scan test and the rank scan test --- are calibrated without any knowledge of $F_0$ and $F_v$.

\subsection{Exponential models} \label{sec:exponential}

Although some of our results will be presented in the general setting above, it is useful to consider an important special case.  This serves as a benchmark we can use to compare the performance of the proposed procedures against that of the optimal tests.  Doing so is classical in the literature on nonparametric tests \citep{MR758442}, where such a test is compared with the likelihood ratio test in some parametric model (often a location model or a scale model).  

In this paper we consider a generic one-parameter exponential model in natural form. Let $F_0$ be a probability distribution on the real line with all the moments finite.  This distribution can be either continuous (i.e., diffuse), discrete (i.e., with discrete support) or a mixture of both.  In the exponential model there is a parameter $\theta_v$ associated with each $v\in\cV$, and the distribution $F_v\equiv F_{\theta_v}$ is defined through its density $f_{\theta_v}$ with respect to $F_0$: for $\theta \in [0, \theta_\star)$, define $f_\theta(x) = \exp(\theta x - \log \varphi_0(\theta))$, where $\varphi_0(\theta) = \int e^{\theta x} {\rm d}F_0(x)$ and $\theta_\star = \sup\{\theta > 0: \varphi_0(\theta) < \infty\}$, assumed to be strictly positive (and possibly infinite). 
In other words, $f_{\theta_v}$ denotes the Radon-Nykodym derivative of $F_{\theta_v}$ with respect to $F_0$.
%For convenience we use the notation $F_v$ and $F_{\theta_v}$ interchangeably.%
Since a natural exponential family has the monotone likelihood ratio property\footnote{A family of densities $(f_\theta : \theta \in \Theta)$, where $\Theta \subset \bbR$, has the monotone likelihood ratio property if $f_{\theta'}(x)/f_\theta(x)$ is increasing in $x$ when $\theta' > \theta$.}, it follows that $F_\theta$ is stochastically increasing in $\theta$ \cite[Lem 3.4.2]{MR2135927}.  In particular, we do have $F_\theta \succeq F_0$ for all $\theta > 0$.  
Important special cases of such an exponential model include the normal location model --- with $F_\theta$ corresponding to $\cN(\theta, 1)$ --- standard in many signal and image processing applications; the Poisson model --- with $F_\theta$ corresponding a Poisson distribution --- popular in syndromic surveillance \citep{kulldorff2005stp}; and the Bernoulli model \citep{walther2010optimal} with $F_\theta$ corresponding to a Bernoulli distribution.

Note that in the formulation above the alternative hypothesis is composite.  Tackling this problem using a generalized likelihood ratio approach is popular in practice \citep{Kul} and often referred to as the scan test, as it works by scanning over the possible anomalous sets to determine if there is such a set that is able to ``explain'' the observed data.
Assuming the nonzero $\theta_v$'s are all equal to $\theta$ under the alternative, and that all subsets in the class $\bbS$ have same size, some simplifications lead to considering the test that rejects for large values of the scan statistic
\beq \label{scan0}
\max_{\cS \in \bbS} \sum_{v \in \cS} X_v\ .
\eeq
When the subsets in the class $\bbS$ may have different sizes, a more reasonable approach includes a normalization of the partial sums above, leading to the following variant of the scan statistic
\beq \label{scan}
\max_{\cS \in \bbS} \frac1{\sqrt{|\cS|}} \sum_{v \in \cS} (X_v - \E_0(X_v))\ .
\eeq
($\E_\theta$ denotes the expectation with respect to $F_\theta$, and for a discrete set $\cS$, $|\cS|$ denotes its cardinality.)
As argued in \citep{arias2013cluster}, this test is in a certain sense asymptotically equivalent to the generalized likelihood ratio test.

\subsection{Calibration by permutation} \label{sec:permutation-intro}

Suppose we are considering a test that rejects the null for large values of a test statistic $T(\bX)$ where $\bX = (X_v, v \in \cV)$.  Let $\bx = (x_v, v \in \cV)$ the observed value of $\bX$.  If we were to know the null distribution $F_0$, we would return the P-value as $\P_0(T(\bX) \ge T(\bx))$.
In practice, even with the knowledge of $F_0$ computing the exact P-value might be difficult, but one can approximate it to an arbitrary accuracy and estimate it by Monte Carlo simulation.

Ignoring computational constraints for the moment, calibration by permutation amounts to computing $T(\bx_\pi)$ for all $\pi \in \cV!$, where $\cV!$ denotes the set of all permutations of $\cV$ and $\bx_\pi = (x_{\pi(v)}, v \in \cV)$ is the permuted data.  We then return the P-value 
\[
\frac1{|\cV|!} \left|\left\{\pi \in \cV! : T(\bx_\pi) \ge T(\bx)\right\}\right|
\]
and the rejection decision is based on this value.  Let $M=|\{T(\bx_\pi):\pi\in\cV!\}|$. If there are no multiplicities, meaning $M=\cV!$, it can be shown such tests are exact and that under the null the P-value has a (discrete) uniform distribution on $\{1/M,2/M,\ldots,1\}$. Otherwise the test will be slightly conservative \citep{MR2135927}.  In practice, the number of permutations is very large (as $|\cV !| = |\cV|!$) and the P-value is estimated by simulation (by uniform sampling of permutations).

In our setting, $T$ above will be a form of a scan statistic, similar to the one in \eqref{scan}, which maximizes a standardized sum of data entries over a class $\bbS$ of possible anomalous sets. When calibrating by permutation we are comparing the value $T(\bx)$ of this statistic on the original data $\bx$ with the corresponding value $T(\bx_\pi)$ on permuted data $\bx_\pi$. This is only sensible if the class $\bbS$ has some structure, and in particular it cannot be invariant under permutations. In this paper we consider what is perhaps the simplest such class, which is the class of intervals
\[\cV = \{1,\dots,N\} \text{ and } \bbS = \big\{\{a, \dots, b\} : 1 \le a \leq b \le N \big\}\ .\]
In the next section we elaborate on other possible structural constrains, and the theoretical approach we develop can be used to study the calibration by permutation in those settings as well.

Assuming $T$ has been chosen, we define the oracle scan test as the scan test calibrated with full knowledge of the null distribution by Monte Carlo simulation, and the permutation scan test as the scan test calibrated by permutation as explained above.

\medskip\noindent
{\bf Contribution 1:} {\em We characterize the performance of the permutation scan test in the context of the exponential family, concluding that it has as much asymptotic power as the oracle scan test (Theorem~\ref{thm:perm_test}).}
\medskip

We note that permutation tests are known to perform this well in classical two-sample testing \citep{MR2135927}.  
However, in the context of the scan test, we are only aware of one other paper, that of \cite{walther2010optimal}, that develops theory for the permutation scan test.  This is done in the context of binary data (a Bernoulli model).  Our analysis extends the theory to any natural exponential model as described in \secref{exponential} (which also includes the binary case).  This requires a different set of tools.

\subsection{Scanning the ranks}
As explained earlier, when calibrating by permutation the computation of the scan statistic $T$ must be done for a large enough number of permutations of the original dataset.  Even though this is done for only a relatively small number of permutations, that number is often chosen in the hundreds, if not thousands, meaning that the procedure requires the computation of that many scans. Even if the computation (in fact, approximation) of the scan statistic is done in linear time this can be rather time consuming.  Furthermore, for a new instantiation of the data the whole procedure must be undertaken anew.  The computational burden of doing so may be prohibitive in some practical situations, for instance, when monitoring a sensor network in real-time.

To mitigate those drawbacks we propose instead a rank-based approach, which avoids the expensive calibration by permutation.
The procedure amounts to simply replacing the observations with their ranks\footnote{Throughout, the observations are ranked in increasing order of magnitude.} before scanning, so that we end up scanning the ranks instead of the original values.  If ties in the ranks are broken randomly the resulting test statistic is distribution-free and therefore can be calibrated by Monte Carlo simulation requiring only the knowledge of the data size (which is $N\equiv|\cV|$ in our context). 
In terms of computational complexity this procedure is as complex as the implementation of a scan test when the null distribution is fully known so there is no computational disadvantage in using ranks.  In fact faster implementations might be possible by taking advantage of the discrete nature of the ranks and avoiding floating-point algebra, but these algorithmic considerations are beyond the scope of this paper.

\medskip\noindent
{\bf Contribution 2:} {\em We establish the performance of the rank scan test (Theorem~\ref{thm:rank_test} and and Proposition~\ref{prp:small}). In the context of the exponential family we show that it has nearly as much asymptotic power as the oracle scan test (Proposition~\ref{prp:rank_test_exp}).}
\medskip

This result is remarkable in the sense that the scan test can be completely calibrated before any data has been observed, and yet attain essentially the same power as the optimal test with full knowledge of the statistical model.  Such a procedure is very natural (albeit distinct) given the classical literature on nonparametric tests \citep{MR758442}, and rank tests such as Wilcoxon's are known to perform this well in classical two-sample testing \citep{MR758442,MR2135927}.

Our results allow us to precisely quantify how much (asymptotic) power is lost when using the rank scan test versus the oracle scan test.  
For example, in the normal means model the rank scan test requires a signal magnitude 1.023 times larger than the regular scan test to be asymptotically powerful against anomalous sets that are not too small.

\subsection{Structured anomalies} \label{sec:intervals-intro}

Naturally, the intrinsic difficulty of the detection task depends not only on the data distribution, but also on the complexity of the class of anomalous sets $\bbS$. Furthermore, for the permutation or rank-based approaches to be sensible this class must have some structure and not be invariant under permutations, as seen above.  In several scenarios structural assumptions on such classes arise very naturally. For instance, grid-like networks are an important special case, arising in applications such as signal and image processing (where the signals are typically regularly sampled) and sensor networks deployed for the monitoring of some geographical area, for example.
This situation is considered in great generality and from different perspectives in \citep{arias2011detection,walther2010optimal,MGD,morel,perone,cai2014rate,HJ09}.
Also, the distribution of the corresponding scan statistic \eqref{scan} and variants has been studied in a number of places \citep{jiang,boutsikas,sieg95,kabluchko2011extremes,sharpnack2014exact}.

The simplest and most emblematic setting is that of detecting an interval in a one-dimensional regularly sampled signal, that was highlighted above. However, the principles underlying the detection of intervals can be used for the detection of much more general anomaly classes. As shown in \citep{arias2011detection}, similar results apply to a general (nonparametric) class $\bbS$ of blob-like (`thick') sets $\cS$ when $\cV$ is a grid-like set of arbitrary finite dimension, although the scanning is done over an appropriate approximating net for $\bbS$ (instead of the entire class $\bbS$). 
Furthermore, these results generalize to one-parameter exponential models, beyond the commonly assumed normal location model, as long as the sets $\cS \in \bbS$ are sufficiently large (poly-logarithmic in $N$). Other papers that develop theory for different environments include \citep{sharpnack2010identifying,sharpnack2013near,maze,addario2010combinatorial,zhao2009anomaly}.  Variants of this detection problem have been suggested, and the applied literature is quite extensive.  We refer the reader to \citep{arias2011detection} and references therein.

Since the main motivation of our work is to develop methods and theory for the scenario when the distributions are unknown/unspecified we focus exclusively on the detection of intervals, for the sake of clarity and simplicity. Nevertheless our techniques and results apply naturally to more general anomaly classes (e.g., rectangles in two or more dimensions, or even blob-like subsets).  The key to these generalizations are proper concentration inequalities for sampling without replacement, namely Lemmas~\ref{lem:bernstein} and  \ref{lem:chernoff}, and a geometric characterization of the anomaly class in terms of an approximating net akin to Lemma~\ref{lem:approximating_net}.  The latter characterization is heavily dependent on the class of anomalous sets under consideration, as described in the preceding paragraph.
Furthermore, although it is possible to study a version of the test than scans over all possible anomalous sets, we choose to study a scan test restricted to an approximating net because of the following advantages: the analysis is simpler as it does not require the use of chaining to achieve tight constants; it is applicable in more general settings, in particular when the class $\bbS$ is nonparametric; it is computationally advantageous as it gives rise to fast implementations.

\subsection{Content and notation}

The rest of the paper is organized as follows.
In \secref{known} we consider the case when the null distribution is known.  This section is expository, introducing the reader to the basic proof techniques that are used, for example, in \citep{arias2011detection}, to establish the performance of the scan statistic when calibrated with full knowledge of the null distribution --- the oracle scan test, as we called it here.  To keep the exposition simple, and to avoid repeating the substantially more complex arguments detailed in that paper and others, we focus on the problem of detecting an interval in a one-dimensional lattice.
This allows us to set the foundation and discover what the performance bounds for the scan test in this case rely on.  
In \secref{permutation} we consider the same setting and instead calibrate the scan statistic by permutation.
In \secref{ranks} we consider the same setting and instead scan the ranks.
In both cases, our analysis relies on concentration inequalities for sums of random variables obtained from sampling without replacement from a finite set of reals, already established in the seminal paper of \cite{hoeffding}.  
In \secref{numerics} we perform some simulations to numerically quantify how much is lost in finite samples when calibrating by permutation or when using ranks.
We also compare our methodology with the method of \cite{tony2012robust}, on simulated data, and also on a real dataset from genomics.
\secref{discussion} is a brief discussion.
Except for the expository derivations in \secref{known}, the technical arguments are gathered in \secref{proofs}.

\section{When the null distribution is known} \label{sec:known}

This section is meant to introduce the reader to the techniques underlying the performance bounds developed in \citep{arias2011detection,MGD} for the scan statistic (and variants) when the null distribution is known. These provide a stepping stone for our results in regards to permutation and rank scan tests. We detail the setting of detecting an interval of unknown length in a one-dimensional lattice.  Therefore, as in \secref{permutation-intro}, consider the setting where 
\[\cV = \{1,\dots,N\} \text{ and } \bbS = \big\{\{a, \dots, b\} : 1 \le a \leq b \le N \big\}\ .\] 
We begin by considering the normal model --- $X_v \sim \cN(\theta_v, 1)$ are independent --- and explain later on how to generalize the arguments to an arbitrary exponential model as described in \secref{exponential}. 
We are interested in testing 
\beq\label{problem}
H_0: \theta_v = 0, \forall v \in \cV \quad \text{ versus }\quad H_1: \exists \cS \in \bbS :\ \tfrac{1}{|\cS |} \sum_{v \in \cS} \theta_v \ge \tau \sqrt{2 \log (N)/|\cS |}\ ,
\eeq
where $\tau>0$ is fixed.  We consider this problem from a minimax perspective.
It is shown in \citep{MGD} that, if $\tau < 1$, then any test with level $\alpha$ has power at most $\beta(\alpha, N)$, with $\beta(\alpha, N) \to \alpha$ as $N \to \infty$.
In other words, in the large-sample limit, no test can do better than random guessing --- the test that rejects with probability $\alpha$ regardless of the data.
On the other hand, if $\tau>1$, then for any level $\alpha>0$ there exists a test with level $\alpha$ and power $\beta(\alpha, N)\to 1$ as $N\to\infty$. In particular, such a test can be constructed using a form of scanning over an approximating net, as explained in the rest of this section.

\medskip\noindent {\em Step 1: Construction of an approximating net.}
Instead of scanning over $\bbS$ we will scan over a subclass of intervals $\bbS_b$, where $0\leq b \leq N$ is an integer to be specified later on.  This brings both computational and analytical advantages over scanning all sets in $\bbS$ as discussed in \secref{intervals-intro}.  Such a subclass must satisfy two important properties, namely have cardinality significantly smaller than $\bbS$, and be such that any element $\cS\in\bbS$ can be ``well approximated'' by an element of $\cS^* \in\bbS_b$.  By well approximated we mean that $\rho(\cS,\cS^* )\approx 1$ where
$$\rho(\cS,\cS^*) := \frac{|\cS \cap \cS^* |}{\sqrt{|\cS| |\cS^* |}}\ ,$$
is a measure of similarity of two sets.  We use an approximating net similar to that of \citep{MGD}; see \citep{sharpnack2014exact} for an alternative construction.

To simplify the presentation assume $N$ is a power of 2 (namely $N=2^q$ for some integer $q$).  Let $\bbD_j$ denote the class of dyadic intervals at scale $j$, meaning of the form $\cS = [1+ k 2^j, (k+1) 2^j] \subset \cV$ with $j$ and $k$ nonnegative integers.  
Let $\bbD_{j,0}$ denote the class of intervals of the form $S \cup S'$ with $S, S' \in \bbD_{j-1}$. Note that $\bbD_j \subset \bbD_{j,0}$.
Then, for $1\leq k < b$, let $\bbD_{j,k}$ be the class of intervals of $\cV$ of the form $S_{\rm left} \cup S \cup S_{\rm right}$, where $S \in \bbD_{j,k-1}$ while $S_{\rm left}$ (resp.~$S_{\rm right}$) is adjacent to $S$ on the left (resp.~right) and is either empty or in $\bbD_{j-k}$. Note that $\bbD_{j,k-1} \subset \bbD_{j,k}$ by construction.
In the last step, $\bbD_{j,b}$ is of the same form as before, only the appended intervals $S_{\rm left}$ and $S_{\rm right}$ are either empty, or in $\bbD_{j-b+1}$.
Finally, define $\bbS_b = \bigcup_j \bbD_{j,b}$.

We can prove the following result for this approximating net, using similar arguments to those of \cite{MGD}.

\begin{lem}\label{lem:approximating_net}
The subclass $\bbS_b \subset \bbS$ has cardinality at most $N 4^{b+1}$ and is such that for any element $\cS\in\bbS$ there is an element $\cS^*\in\bbS_b$ satisfying $\cS\subset \cS^*$ and $\rho(\cS,\cS^*) \ge (1 + 2^{-b+2})^{-1/2}$.
\end{lem}

\begin{rem}\label{rem:computational_complexity}
It is easy to see that the subclass $\bbS_b$ can be scanned in $O(N b 4^b)$ operations --- this is implicit in \citep{MGD}.  
Indeed, we start by observing that scanning all dyadic intervals can be done in $O(N)$ operations by recursion, starting from the smallest intervals and moving up (in scale) to larger intervals.  We then conclude by realizing that each interval in $\bbS_b$ is the union of at most $2b+2$ dyadic intervals. 
\end{rem}

\medskip\noindent {\em Step 2: Definition of the scan test.}
We consider a test based on scanning only the intervals in $\bbS_b$. This test rejects the null if
\beq \label{scan-net}
\max_{\cS \in \bbS_b} Y_\cS \ge \sqrt{2(1+\eta) \log N}\quad\text{ with }\quad Y_\cS := \frac1{\sqrt{|\cS|}} \sum_{v \in \cS} X_v\ ,
\eeq
where $\eta >0$ satisfies $\eta \to 0$ and $\eta \log(N) \to \infty$.  (The reason for these conditions will become clear shortly.) 

\medskip\noindent {\em Step 3: Under the null hypothesis.}
By the union bound, we have
\[\begin{split}
\P_0\left(\max_{\cS \in \bbS_b} Y_\cS \ge \sqrt{2(1+\eta) \log N}\right) 
&\le \sum_{\cS \in \bbS_b} \P_0\left(Y_\cS \ge \sqrt{2(1+\eta) \log N} \right) \\
&\le |\bbS_b| \bar\Phi\left(\sqrt{2(1+\eta) \log N}\right)\ , 
\end{split}\]
where $\Phi$ denotes the standard normal distribution function and $\bar \Phi = 1 - \Phi$ denotes the corresponding survival function. We have the well-known bound on Mill's ratio:
\beq \label{mills}
\bar \Phi(x) \le e^{-x^2/2}, \quad \forall x \ge 0\ .
\eeq
Therefore we get
\[
\P_0\left(\max_{\cS \in \bbS_b} Y_\cS \ge \sqrt{2(1+\eta) \log N}\right) 
\le N 4^{b+1} N^{-(1+\eta)}
= N^{-\eta} 4^{b+1}\ .
\]
We choose $b = \frac12 \eta \log(N)/\log(4)$.  With our assumption that $\eta \log N \to \infty$, this makes the last expression tend to zero as $N \to \infty$.  (It also implies that $b \to \infty$, which we use later on.)
We conclude the test in \eqref{scan-net} has level tending to 0 as $N\to \infty$.  

\medskip\noindent {\em Step 4: Under the alternative.}
We now show that the power of this test tends to 1 when $\tau > 1$.
Let $\cS$ denote the anomalous interval.
Referring to Lemma~\ref{lem:approximating_net}, there is a set $\cS^*\in\bbS_b$ such that $\rho(\cS,\cS^*) \geq (1 + 2^{-b+2})^{-1/2}$, so that $\rho(\cS,\cS^*) = 1 +o(1)$ since $b \to \infty$.
Furthermore $Y_{\cS^*}$ is normal with mean at least $\rho(\cS,\cS^*) \tau \sqrt{2 \log N}$ and variance 1.  
We thus have
\[
\P\left(Y_{\cS^*} \ge \sqrt{2(1+\eta) \log N}\right) \geq \bar\Phi(\xi)\ ,
\]
where
\begin{align*}
\xi &:= \sqrt{2(1+\eta) \log N} - \rho(\cS,\cS^*) \tau \sqrt{2 \log N}\\
&= \sqrt{2(1+\eta) \log N}\big(1 - (1+o(1)) \tau/\sqrt{1+\eta}\big)\\
&\sim - (\tau-1) \sqrt{2 \log N} \to -\infty\ ,
\end{align*}
where we used the fact that $\tau>1$ is fixed and $\eta\to 0$.  We conclude that the test in \eqref{scan-net} has power tending to 1 as $N\to \infty$. In conclusion, we have shown the following result.

\begin{prp}[\cite{MGD}] \label{prp:MGD}
Refer to the hypothesis testing problem in \eqref{problem}. The test defined in \eqref{scan-net}, with $\eta = \eta_N \to 0, \eta_N \log N \to \infty$ and $b = b_N = \frac12 \eta_N \log N$, has level converging of 0 as $N\to\infty$.  Moreover when $\tau > 1$ it has power converging to 1 as $N\to\infty$.
\end{prp}

We remark that, in principle, we may choose any $b = b_N \to \infty$ such that $b_N/\log N \to 0$. From Remark~\ref{rem:computational_complexity} the computational complexity of the resulting scan test is of order $O(N b_N 4^{b_N})$.  For example, $b_N \sim \log\log N$ is a valid choice and the resulting scan test runs in $O(N {\rm polylog}(N))$ time.

\subsection{Generalizations} \label{sec:general}

The arguments just given for the setting of detecting an anomalous interval under a normal location model can be generalized to the problem of detecting other classes of subsets under other kinds of distributional models.  We briefly explain how this is done.  
(Note that these generalizations can be combined.)

\paragraph{Other classes of anomalous subsets}
For a given detection problem, specified by a set of nodes $\cV$ and a class of subsets $\bbS \subset 2^\cV$, the arguments above continue to apply if one is able to construct an appropriate approximating net as in \lemref{approximating_net}.  This is done, for example, in \citep{arias2011detection,MGD} for a wide range of settings.  We note that the construction of a net is purely geometrical and/or combinatorial.

\paragraph{Other exponential models}
To extend the result to an arbitrary (one-parameter, natural) exponential model, we require the equivalent of the tail-bound \eqref{mills}.  
While such a bound may not apply to a particular exponential model, it does apply asymptotically to large sums of IID variables from that model by Chernoff's bound and a Taylor development of the rate function.

Indeed, recalling the notation introduced in \secref{exponential}, let $\psi_0(t) = \sup_{\lambda\in[0,\theta^*)} (\lambda t - \log \varphi_0(\lambda))$, which is the rate function of $F_0$.  By Chernoff's bound, we have
\beq \label{chernoff}
\P_0(Y_\cS \ge y) \le \exp\Big(-|\cS| \psi_0(y |\cS|^{-1/2})\Big)\ . 
\eeq
Assuming without loss of generality that $F_0$ has zero mean and unit variance, we have 
\beq \label{rate0}
\psi_0(t) \ge \frac12 t^2 + O(t^3)\ , \quad t \to 0\ .
\eeq
To see this, note that $\varphi_0(\lambda)$ is infinitely many times differentiable when $\lambda\in[0,\theta^*)$, with $\varphi_0'(0) = \E_0(X) = 0$ and $\varphi_0''(0) = \E_0(X^2) = 1$.  Therefore $\varphi_0(\lambda) = 1 + \frac12 \lambda^2 +O(\lambda^3)$ as $\lambda\to 0$.
For $t \in [0,\theta^*)$, we then have
\begin{align*}
\psi_0(t) &= \sup_{\lambda \in [0,\theta^*)} \big[\lambda t - \varphi_0(\lambda)\big] \geq t^2 - \log \varphi_0(t)\\
&= t^2 - \log \left( 1+\frac12 t^2+O(t^3) \right) \geq \frac12 t^2+O(t^3)\ ,
\end{align*}
where we use $\log(1+x)\leq x$. From this we see that our derivations for the normal model apply essentially verbatim if, for some constant $c >0$, $|\cS| \ge c (\log N)^3$ for all $\cS \in \bbS$. Furthermore, it can be seen that the test in \eqref{scan-net} is essentially optimal for exponential models, as its performance matches the lower bounds in \citep{arias2011detection}.

%%%%%%%%%%%%%%%%%%%%%%%%%%%%%%%%%%%%%%%%%%%%%%%%%%%%

\section{Calibration by permutation} \label{sec:permutation}
Having described in detail how a performance bound is established for the scan test variant \eqref{scan-net} for the problem of detecting an interval of unknown length, and its extensions to other detection problems, we now clearly see that the key to adapting this analysis to a calibration by permutation is a concentration of measure bound to replace \eqref{mills} and \eqref{chernoff}.
Since this is the same in any detection setting, we consider as in \secref{known} the problem of detecting an interval of unknown length.
This time, we impose a minimum and maximum length on the intervals
\beq \label{bbS-2}
\bbS = \big\{\{a, \dots, b\} : 1 \le a < b \le N, 2^{q_l} \le b-a \le 2^{q_u} \big\}\ .
\eeq
Indeed, when calibrating the scan test by permutation, we necessarily have to assume nontrivial upper and lower bounds on the size of an anomalous interval. To see this consider intervals of length one. Then the value of the scan for any permutation of the data is the same. By symmetry the same reasoning applies for intervals of length $N-1$.

We consider essentially the same form of the scan statistic \eqref{scan} as before, but replace $\E_0(X_v)$ (which we do not have access to) by $\bar X=\frac{1}{N}\sum_{v\in\cV} X_v$ and scan over an approximating net.  We restrict the approximating net to match the class of intervals defined in \eqref{bbS-2} (but still call it $\bbS_b$ for simplicity). Specifically we only keep an element $\cS^* \in \bbS_b$ if there is $S \in \bbS$ such that $\rho(\cS, \cS^*) \ge (1+2^{-b+2})^{-1/2}$. This ensures that the statements in \lemref{approximating_net} still hold, and also that $|\cS^*| \ge 2^{q_l}/(1+2^{-b+2})$ for all $\cS^* \in \bbS_b$.  In detail, with $\bx = (x_v, v \in \cV)$ denoting the observed data, we define

\beq \label{scan-b}
\scan(\bx) = \max_{\cS \in \bbS_b} \left(Y_\cS(\bx)-\sqrt{|\cS|}\bar x\right)\ , \quad Y_\cS(\bx) := \frac1{\sqrt{|\cS|}} \sum_{v \in \cS} x_v\ ,
\eeq
The test rejects the null at significance level $\alpha\in(0,1)$ when
\beq \label{scan_perm}
\mathfrak{P}(\bx) := \frac1{|\cV|!} \Big| \big\{\pi \in \cV! : \scan(\bx_\pi) \ge \scan(\bx)\big\} \Big| \leq \alpha\ ,
\eeq
where $\mathfrak{P}(\bx)$ is the permutation P-value.

\begin{thm} \label{thm:perm_test}
Refer to the hypothesis testing problem in \eqref{problem} and assume $F_0$ has zero mean and variance one.  Consider the test that rejects the null if $\mathfrak{P}(\bX)\leq \alpha$, where $\mathfrak{P}$ is defined in \eqref{scan_perm}, with $b = b_N \to \infty$ and $b_N/\log N \to 0$ at $n\to\infty$. This test has level at most $\alpha$. Furthermore, assume that under the alternative the anomalous set $\cS$ belongs to $\bbS$ defined in \eqref{bbS-2} with $q_l - 3 \log_2 \log N \to +\infty$ and $q_u - \log_2 N \to -\infty$ as $N\to\infty$.  This test has power converging to 1 as $N\to\infty$ when
\[
\frac{1}{|\cS |} \sum_{v \in \cS} \theta_v \ge \tau \sqrt{2 \log (N) /|\cS |}, \quad \text{with } \tau > 1 \text{ fixed,}
\]
provided that either $F_0$ has compact support or $\max_v \theta_v \le \tilde\theta < \theta_\star$ for some fixed $\tilde\theta > 0$.
\end{thm}

The headline here is that a calibration by permutation has as much asymptotic power as a calibration by Monte Carlo with full knowledge of the null distribution (to first-order accuracy).  This is (qualitatively) in line with what is known in classical settings \citep{MR2135927}.
Note that this testing procedure makes no assumptions about $F_0$ or about the existence of an underlying exponential model.

\begin{rem}
The assumption that $F_0$ has zero mean and variance one is without any loss of generality, and merely for clarity of presentation. In general, the permutation-based test is asymptotically powerful under the alternative if there is a set $\cS\in\bbS$ such that
$$\frac{1}{|\cS |} \sum_{v \in \cS} \theta_v \geq \tau\frac{1}{\sigma_0}\sqrt{2\log (N)/|\cS |}, \quad \text{with } \tau > 1 \text{ fixed,}$$
where $\sigma_0^2$ denotes the variance of $F_0$.
\end{rem}

The conditions required here allow $\bbS$ to be any class of intervals of lengths between $(\log N)^{3 + a}$ and $o(N)$, for any $a > 0$ fixed. 
This includes the most interesting cases of intervals not too short and also not too long. 
In fact, for certain families of distributions removing from consideration very small intervals is essential and cannot be avoided. 

\begin{ex} \label{ex:bernoulli}
For instance consider the Bernoulli model, where $X_v \sim \text{Bernoulli}(1/2)$, for all $v \in \cV$ under the null, and $X_v \sim \text{Bernoulli}(1)$, for all $v\in \cS$ when $\cS$ is anomalous.  Even under the null we will encounter a run of ones of length $\sim \log_2 N$ (the famous Erd\H os--R\'enyi Law) with positive probability.  Therefore in this case the scan test, calibrated by Monte Carlo or permutation, is powerless for detection of intervals of length $\frac12 \log_2 N$.  In fact, it can be shown that no test has any power in that case.
\end{ex}

Note that, when calibrating a test by permutation there are essentially two sources of randomness. The randomness intrinsic to the data $\bX$, and the randomness induced by the permutation. In particular, if we regard $\pi$ as a uniform random variable over the set of possible permutations $\cV!$ the P-value of the test can be re-written as $\mathfrak{P}(\bX)=\P\left(\scan(\bX_\pi) \ge \scan(\bX)\right)$. Under the null hypothesis the argument is classic: for any given permutation $\pi$, the distribution of $\bX$ is identical to the distribution of $\bX_\pi$, therefore $\scan(\bX)$ is conditionally uniformly distributed in $\{\scan(\bX_\pi):\pi\in\cV!\}$ (with multiplicities). The bulk of the effort in the proof is to characterize the behavior of the test under the alternative. The first step is to, conditionally on the data $\bX$, ``remove'' the randomness in $\pi$. Realizing that for any $\cS$, $\sum_{v\in\cS} X_\pi(v)$ is simply a sum of elements sampled without replacement from $\bX$, we are able to use a concentration inequality for sampling without replacement to upper-bound the P-value by an expression involving $\scan(\bX)$, the sample mean and variance of $\bX$, and $\max_v X_v$. The remainder of the proof consists in controlling those terms for the exponential model.

For technical reasons, we place an upper bound $\tilde\theta$ on the nonzero $\theta_v$'s to streamline the proof arguments and be able to control $\max_v X_v$.  However, note that this condition is not a simple artifact of the proof technique and its removal will invalidate the statement.  A way around this assumption is to state the result in terms of $\min_{v\in\cS} \theta_v$ instead of $\frac{1}{|\cS|}\sum_{v\in\cS} \theta_v$ and use censoring prior to scanning (see the discussion in \secref{discussion}).

%%%%%%%%%%%%%%%%%%%%%%%%%%%%%%%%%%%%%%%%%%%%%%%%%%%%

\section{Scanning the ranks}\label{sec:ranks}

Having observed $\bx = (x_v, v \in \cV)$, scanning the ranks amounts to replacing every observation with its rank among all the observations, and computing the scan \eqref{scan-b}.  
We call this the {\em rank scan}.
As for all rank-based methods, the null distribution is the permutation distribution when there are no ties.
\bitem
\item When there are no ties with probability one, calibration of the distribution of the test statistic is determined by the data size $N$, and therefore the test can be calibrated by Monte Carlo simulation before data is observed.
\item When there are ties the rank scan test can be also calibrated by permutation. If one breaks ties using the average rank then calibration must be done anew for any given dataset.  A much better alternative is to break ties randomly so that we are back in the first case, and can calibrate the test before seeing the data.  The latter option is computationally superior and is the one we analyze.
\eitem

In summary, the rank scan test is computationally more advantageous, when compared with the test of the previous section, calibrated by permutation.  An additional advantage of the rank scan is its robustness to outliers --- although the permutation scan after censoring (discussed in \secref{discussion}) is also robust to outliers.  See \secref{numerics} for implementation issues and a computational complexity analysis.

Formally, let $\bx = (x_v, v \in \cV)$ denote the observations as before, and for every $v \in \cV$, let $r_v$ be the rank (in increasing order) of $x_v$ in $\bx$, where ties are broken randomly, and let $\br = (r_v, v \in \cV)$ be the vector of ranks. The rank scan test returns the P-value $\mathfrak{P}(\br)$ defined in \eqref{scan_perm}.

%As we mentioned in the Introduction, an important advantage of the rank scan over the permutation scan is the fact that the former only requires calibration once, while the latter requires a new calibration with each dataset.
%This assumes that the size $|\cV|$ of the node set remains the same.

Because the rank scan test is naturally regarded as a kind of permutation scan test, we assume similarly upper and lower bounds on the size of the anomalous set as in \secref{permutation}.  The first result we present is rather general, and it is not particular to the exponential family and applies to the general setting in \secref{general}.  For rank-based procedures the performance will depend naturally on the ability to rank correctly an anomalous observation against a normal one. This is naturally captured by the following quantity:
\beq \label{p_v}
p_v = \P (Y>X) + \tfrac{1}{2} \P (Y=X), \quad \text{where $X\sim F_0$ and $Y\sim F_v$ are independent.}
\eeq
The larger $p_v$ is the higher is the probability of ranking the two observations correctly.

\begin{thm}\label{thm:rank_test}
Refer to the hypothesis testing problem in \secref{general} and consider the test that rejects the null if $\mathfrak{P}(\bR)\leq \alpha$, where $\mathfrak{P}$ is defined in \eqref{scan_perm}, with $b = b_N \to \infty$ and $b_N/\log N \to 0$. This test has level at most $\alpha$. Furthermore this test has power converging to 1 as $N\to\infty$ provided
$$\frac{1}{|\cS |} \sum_{v\in\cS} p_v \geq \frac{1}{2}+ \tau \sqrt{2\log (N)/|\cS |}\ , \quad \text{with } \tau > \frac1{2\sqrt{3}} \text{ fixed,}$$
and $\cS$ belongs to $\bbS$ defined in \eqref{bbS-2} with $q_l - \log_2 \log N \to +\infty$ and $q_u - \log_2 N \to -\infty$ as $N\to\infty$.
\end{thm}

This result characterizes the performance of the rank scan test for general distributions (actually we do not even need to assume $F_v$ stochastically dominates $F_0$). To get a better sense of this result and be able to compare it with the previous theorem it is useful to consider the particular case of the exponential model. Define
\beq \label{Upsilon}
\Upsilon_0 = \frac{1}{2}\E[\max(X,Y)]\ ,
\eeq
where $X,Y \sim F_0$ and independent.
\begin{prp}\label{prp:rank_test_exp}
Refer to the hypothesis testing problem in \eqref{problem}, assume $F_0$ has zero mean and variance one, and refer to the test in Theorem~\ref{thm:rank_test}. The test has level at most $\alpha$.  Moreover, it has power converging to 1 as $N\to\infty$ when
\[
\frac{1}{|\cS |} \sum_{v\in \cS} \theta_v \geq \tau \sqrt{2 \log (N)/|\cS |} \ , \quad \text{with } \tau > \frac{1}{2\sqrt{3} \Upsilon_0} \text{ fixed.}
\]
\end{prp}

The headline here is that the rank scan requires a signal amplitude which is $1/(2\sqrt{3} \Upsilon_0)$ larger than what is required of the regular scan test calibrated by Monte Carlo with full knowledge of the null distribution.  This is (qualitatively) in line with similar results in more classical settings \citep{MR758442}.
For the normal location model, we find that $1/(2\sqrt{3}\Upsilon_0) = \sqrt{\pi/3} \approx 1.023$, so the detection threshold of rank scan is almost the same as that of the regular scan test --- see the Appendix~\ref{sec:C0} for details.  Note that $\Upsilon_0\leq 1/(2\sqrt{3})$ (otherwise this would contradict the known minimax lower bounds) and that equality is attained if and only if $F_0$ is the uniform distribution.\footnote{This is based on a personal communication from Richard J.~Samworth and Tengyao Wang, who got interested in this question after one of the present authors presented this work at Cambridge University.}

\begin{rem}
As in the case of Theorem~\ref{thm:rank_test} the assumption on the moments of $F_0$ are used only for clarity of presentation. In general, the permutation-based test is asymptotically powerful under the alternative if there is a set $\cS\in\bbS$ such that
\[\frac{1}{|\cS |} \sum_{v\in\cS} \theta_v \geq \tau \sqrt{2\log (N)/|\cS |}\ , \quad \text{with } \tau > \frac{\tau}{2\sqrt{3} (\Upsilon_0 - \mu_0 /2)} \text{ fixed,}\]
where $\mu_0$ denotes the mean of $F_0$.
\end{rem}

The proof of Theorem~\ref{thm:rank_test} starts essentially as that of Theorem~\ref{thm:perm_test}. Under the alternative the P-value is upper bounded by an expression involving $\scan(\bR)$. Control of this term is more complicated than that of $\scan(\bX)$ in the previous theorem, since the elements of $\bR$ are not independent, but can be done by controlling the first two moments of $\bR$. For Proposition~\ref{prp:rank_test_exp} we note that for the exponential model one can relate $p_v\equiv p_{\theta_v}$ to $\theta_v$ by a Taylor expansion around zero, concluding the proof.

%%%%%%%%%%%%%%%%%%%%%%%%%%%%%%%%%%%%%%%%%%%%%%%%%%%%

\subsection*{Small and very small intervals}

The conditions of \thmref{rank_test} allow for dealing with intervals of length of order (strictly) larger than $\log N$. We give here results that encompass the scenario where the interval might be of smaller length. To keep the discussion simple we consider the class of intervals of a fixed size $|\cS|=k$ under the alternative, and explain later how this result is generalized for a class of intervals of different sizes. In this situation there is no need to consider an approximating net and we simply scan over the entire class, denoted by $\bbS$.
Recall the definition of the permutation P-value \eqref{scan_perm}. 

\begin{prp} \label{prp:small}
Refer to the hypothesis testing problem in \secref{general} and consider the test that rejects the null if $\mathfrak{P}(\bR)\leq \alpha$.  Then the test has level at most $\alpha$ and power converging to 1 as $N\to\infty$ provided there is an interval $\cS$ of length $k$ such that
\begin{itemize}[nolistsep]
\item[(i)] $\frac{1}{|\cS |} \sum_{v\in\cS} p_v = 1- o(N^{-2/k})$ when $2<k=o(\log N)$; or 
\item[(ii)] $\frac{1}{|\cS |} \sum_{v\in\cS} p_v > 1- \frac12 \exp(-\tfrac{c+1}{c})$ when $k = c \log N$ for some $c>0$ fixed.
\end{itemize}
\end{prp}

\thmref{rank_test} and \prpref{small} together cover essentially all interval sizes which are $o(N)$. 
\thmref{rank_test} covers the case of larger intervals, in which case $\frac{1}{|\cS |} \sum_{v\in\cS} p_v$ can go to 1/2 provided it does not converge too fast, and the test is still powerful asymptotically.
In \prpref{small}, a sufficient condition for an asymptotically powerful test is that $\frac{1}{|\cS |} \sum_{v\in\cS} p_v$ goes to 1 at a certain rate when the size of the anomalous interval is $o(\log N)$. If the interval size is $c \log N$ with $c>0$ arbitrary the rank test is asymptotically powerful when $\frac{1}{|\cS |} \sum_{v\in\cS} p_v$ is greater than a constant (strictly larger than $1/2$) depending on $c$. 

Extending this result to the exponential model is not possible without additional knowledge of the family of distributions, as having $p_{\theta_v}$ bounded away from $1/2$ implies $\theta_v$ is bounded away from 0.  As an example, consider the normal means model when $k=o(\log N)$. In this case, we have
\[
p_\theta = \Phi (-\theta/\sqrt{2}) \geq 1- \frac{1}{2} e^{- \theta^2 /4}\ .
\]
Hence, whenever $\frac{1}{2} e^{- {\theta}^2 /4} =o(N^{-2/k})$, the condition in the proposition is met.  This is satisfied when 
\beq \label{theta-small-cond}
\theta = \tau \sqrt{2 \log(N)/k}, \quad \text{with $\tau > 2$ fixed}.
\eeq
This means that in this case the rank scan requires an amplitude at most two times larger than the regular scan test calibrated with full knowledge of the null distribution.

Finally note that the condition $\frac{1}{|\cS |} \sum_{v\in\cS} p_v \to 1$ or $\frac{1}{|\cS |} \sum_{v\in\cS} p_v > 1-\tfrac12 \exp(-\tfrac{c+1}{c})$ might not be possible to meet for certain distributions of the exponential family. 
For instance, in Example~\ref{ex:bernoulli}, $\frac{1}{|\cS |} \sum_{v\in\cS} p_v = 3/4$, a case not covered by \prpref{small} when the interval size is smaller than $c \log N$ and $c$ is small enough. But this is expected since no test has any power if $c$ is sufficiently small.

\begin{rem}
Proposition~\ref{prp:small} considered the case when the size of the anomalous interval is known. However, we could consider the class of intervals of length greater than 2 and at most $\tilde k$ for some given $\tilde k = O(\log N)$. 
In this case we would simply scan the ranks for every fixed interval size up to $\tilde k$ and apply a Bonferroni correction to the P-values.
Following through the steps of the proof, one can see that the rank scan test would be asymptotically powerful when

\smallskip
\begin{itemize}[nolistsep]
\item[(i')] $\frac{1}{|\cS |} \sum_{v\in\cS} p_v = 1- o(N \log N)^{-2/|\cS|}$ when $2<|\cS|=o(\log N)$; or 
\item[(ii')] $\frac{1}{|\cS |} \sum_{v\in\cS} p_v > 1-\frac12\exp(-\tfrac{c+1}{c})$ when $|\cS| = c \log N$ for some $c>0$ fixed.
\end{itemize}
For the normal location model and considering $\tilde k = o(\log N)$, we can see that this is satisfied when \eqref{theta-small-cond} holds.
\end{rem}

\section{Numerical experiments} \label{sec:numerics}

\subsection{Computational complexity}
We already cited some works where fast (typically approximate) algorithms for scanning various classes of subsets are proposed \citep{neill2004rapid,neill2012fast,MGD,walther2010optimal}. 
For example, as we saw in \lemref{approximating_net}, \cite{MGD} design an approximating net $\bbS_b$ for the class of all intervals $\bbS$ that can be scanned in $O(N b 4^b)$. Furthermore, we saw in \prpref{MGD} that this procedure achieves the optimal asymptotic power as long as $b = b_N \to \infty$. 
For example, if $b_N \asymp \log\log N$, then the computational complexity is of order $(N {\rm polylog}(N))$.
 
In any case, suppose that a scanning algorithm has been chosen and let $\cC_N$ denote its computational complexity. 
The oracle scan test and the rank scan test are then comparable, in that they estimate the null distribution of their respective test statistic by simulation, and this is done only once for each data size $N$.  
With this preprocessing already done, the computational complexity of these two procedures is $\cC_N$, the cost of a single scan when applied to data of size $N$.
In contrast, the permutation scan test is much more demanding, in that it requires scanning each of the permuted datasets, and this is done every time the test is applied.
Assuming $B$ permutations are sampled at random for calibration purposes, the computational complexity is $B \cC_N$, that is, $B$ times that of the oracle or rank variants (not accounting for preprocessing).  $B$ is typically chosen in the hundreds ($B=200$ in our experiments), if not thousands, so the computational burden can be much higher for the permutation test.

\subsection{Simulations}
We present the results of some basic numerical experiments that we performed to corroborate our theoretical findings in finite samples.  
We generated the data from the normal location model --- where $F_\theta = \cN(\theta, 1)$ --- which is arguably the most emblematic one-parameter exponential family and a popular model in signal and image processing.  We used the regular scan test, calibrated with full knowledge of the null distribution, as a benchmark.  The permutation scan test and rank scan test were calibrated by permutation.

The test statistic that we use in our experiments is the scan over all intervals of dyadic length.  This subclass of intervals is morally similar to $\bbS_0$ (corresponding to $b=0$) but somewhat richer.
This choice allows us to both streamline the implementation and make the computations very fast via one application of the Fast Fourier Transform per dyadic length.
In detail, letting $\bbS$ denote the class of all discrete intervals in $\cV$, this amounts to taking as approximating set
\[
\bbS_{\rm dyad} = \Big\{\cS \in \bbS : |\cS| = 2^j \text{ for some } j \in \bbN\Big\}.
\]

As explained earlier, the calibration by permutation and the rank-based approach are valid no matter what subclass of intervals is chosen, and in fact, the same mathematical results apply as long as the subclass is an appropriate approximating net.  We encourage the reader to experiment with his/her favorite scanning implementation.

It is easy to see that, for each $\cS \in \bbS$, there is $\cS^* \in  \bbS_{\rm dyad}$ with $\cS^* \subset \cS$ and $|\cS^*| > |\cS|/2$.  Hence,
\[
\min_{\cS \in \bbS} \max_{\cS^* \in \bbS_{\rm dyad}} \rho(\cS, \cS^*) \ge1/\sqrt{2}.
\]
A priori, this implies that scanning over $\bbS_{\rm dyad}$ requires an amplitude $\sqrt{2}$ larger to achieve the same (asymptotic) performance as scanning over $\bbS$ or a finer approximating set as considered previously.   
To simplify things, however, in our simulations we took an anomalous interval of dyadic length, so that the detection threshold is in fact the same as before.

\begin{figure}[h!]
\label{fig:power}
\centering
\includegraphics[scale=0.4]{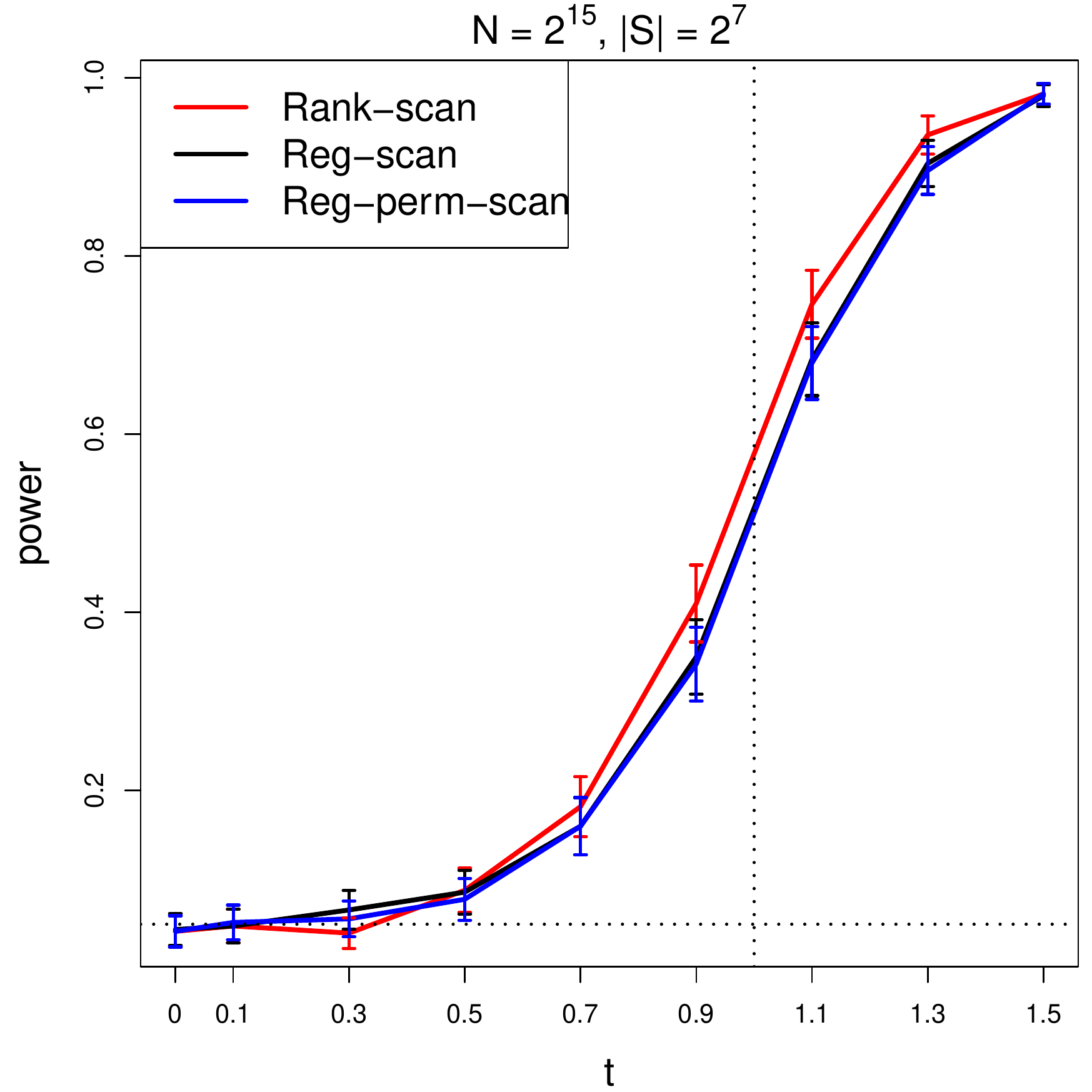} \quad
\includegraphics[scale=0.4]{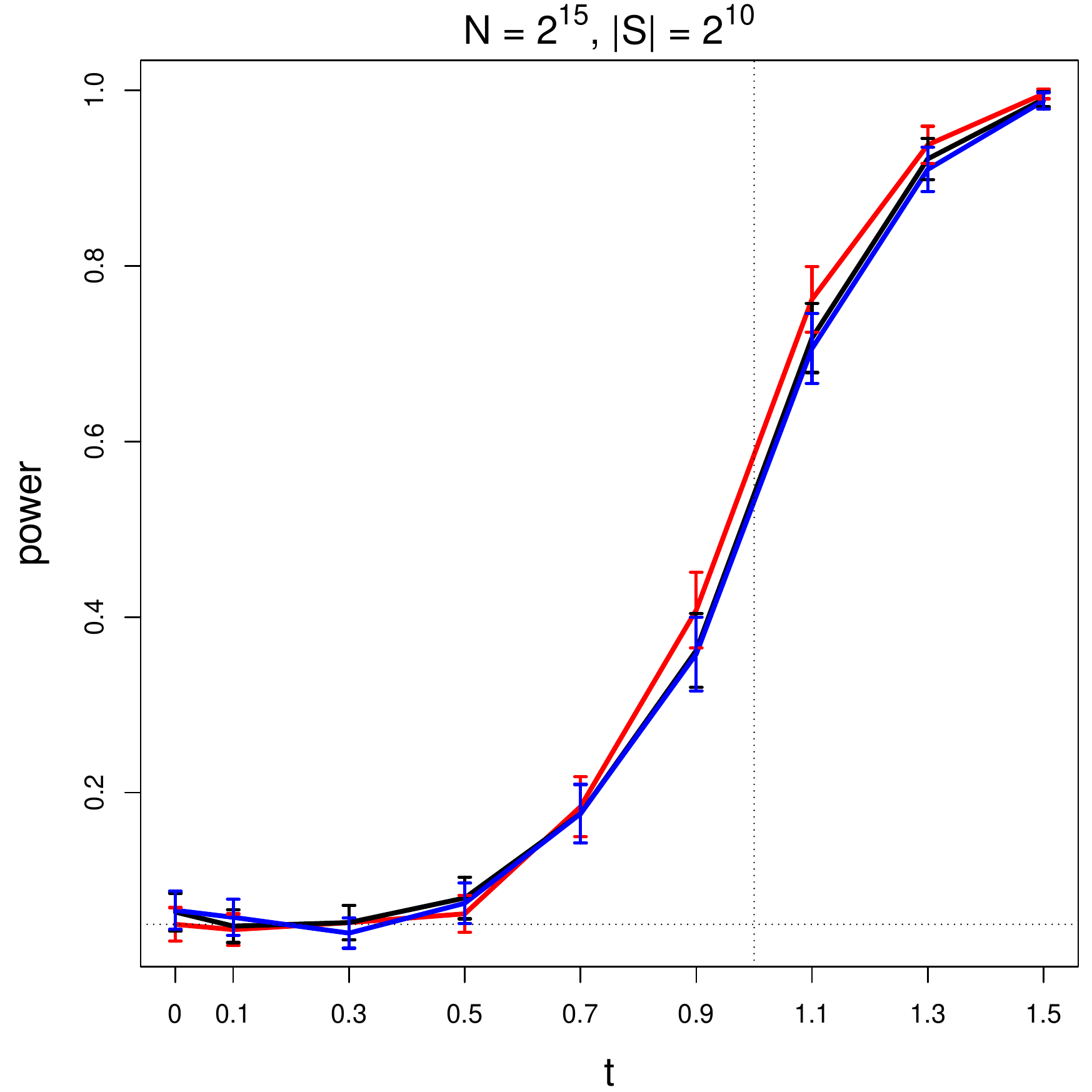}
\caption{Power curves (with 95\% margin of error) for the three tests (all set at level 0.05) as a function of the parameter $t$ in \eqref{t}: the scan test calibrated with knowledge of the null distribution (black); the permutation scan test (blue); and the rank scan test (red).  On the left are the results for $|\cS| = 2^7$ and on the right for $|\cS| = 2^{10}$.  $N = 2^{15}$ in both cases.  Each situation was repeated 1,000 times and each time 200 permutations were drawn for calibration. The vertical black dashed line is the minimax boundary for $t$. The horizontal black dashed line is the significance level $0.05$.}
\end{figure}

We set $N=2^{15}$ and tried two different lengths for the anomalous interval $|\cS| \in \{2^7, 2^{10}\}$.  All the nonzero $\theta_v$'s were taken to be equal to 
\beq \label{t}
\theta_\cS = t \sqrt{2 \log(N)/|\cS|}
\eeq
with $t$ varying.  The critical values and power are based on 1000 repeats in each case.  A level of significance of 0.05 was used.  Also, 200 permutations were used for the permutation scan test.
The results are presented in \figref{power}.  At least in these small numerical experiments, the three tests behave comparably, with the rank scan slightly dominating the others. Although the last finding is somewhat surprising, this is a finite-sample effect and is localized in the intermediate power range (around a power of 0.5) and so does not contradict the theory developed earlier.  In fact, the three tests achieve power 1 at roughly the same signal amplitude, confirming the theory.

\subsection{Comparison with RSI}
Next, we compare our rank scan with the robust segment identifier (RSI) of \cite{tony2012robust}.  This is a recent method based taking the median over bins of a certain size (a tuning parameter of the method) and then scanning over intervals.  Because the median is asymptotically normal, it allows for a calibration that only requires the value of the null density at 0.  In turn, one can try to estimate this parameter.  Although the method is not distribution-free proper, it appears to be the main contender in the literature.  
We first compare the two methods on simulated data, for in the context of detection (the problem we considered so far) and in the context of identification (a problem considered in that paper).

\begin{figure}[h!]
\centering
\includegraphics[scale=0.4]{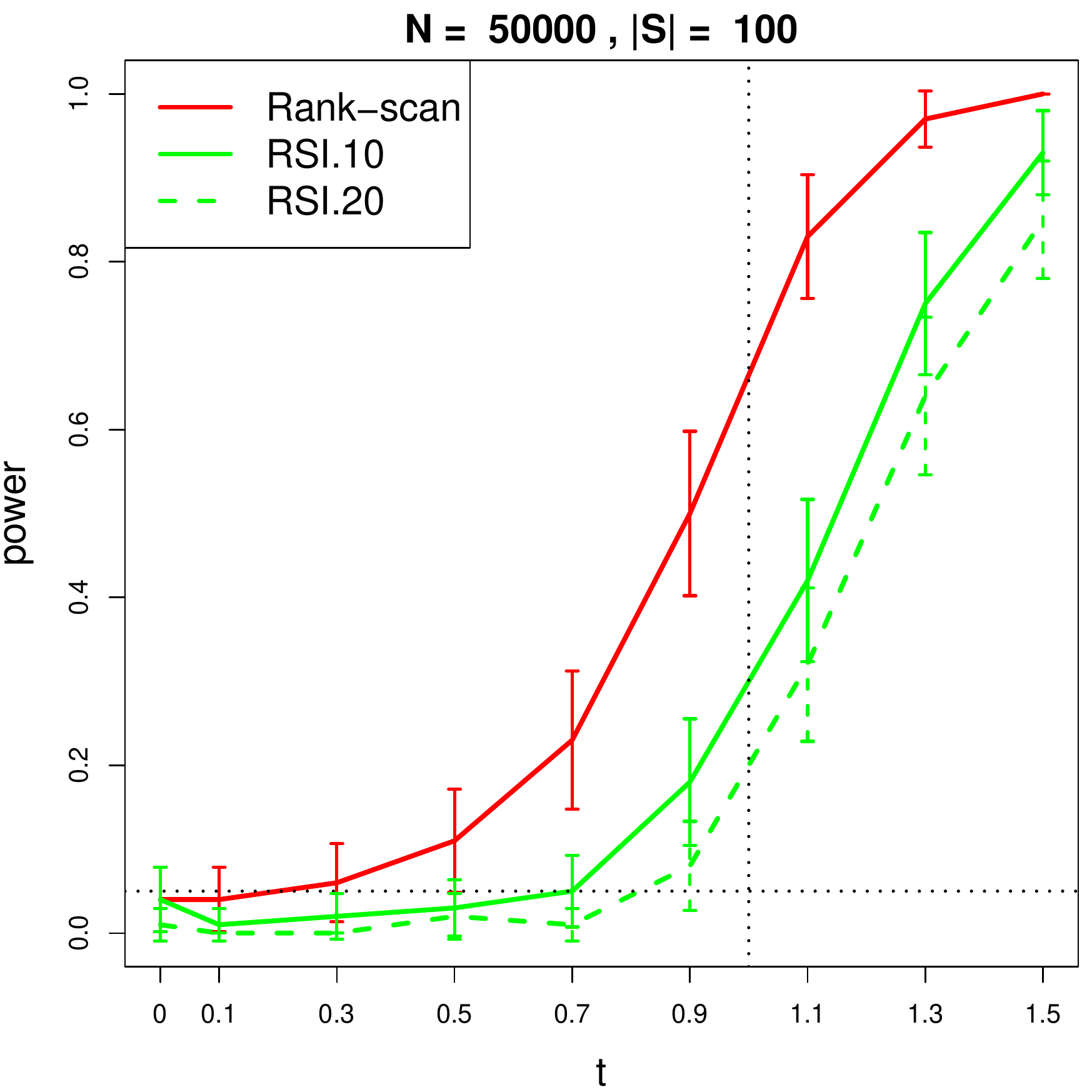} \quad
\includegraphics[scale=0.4]{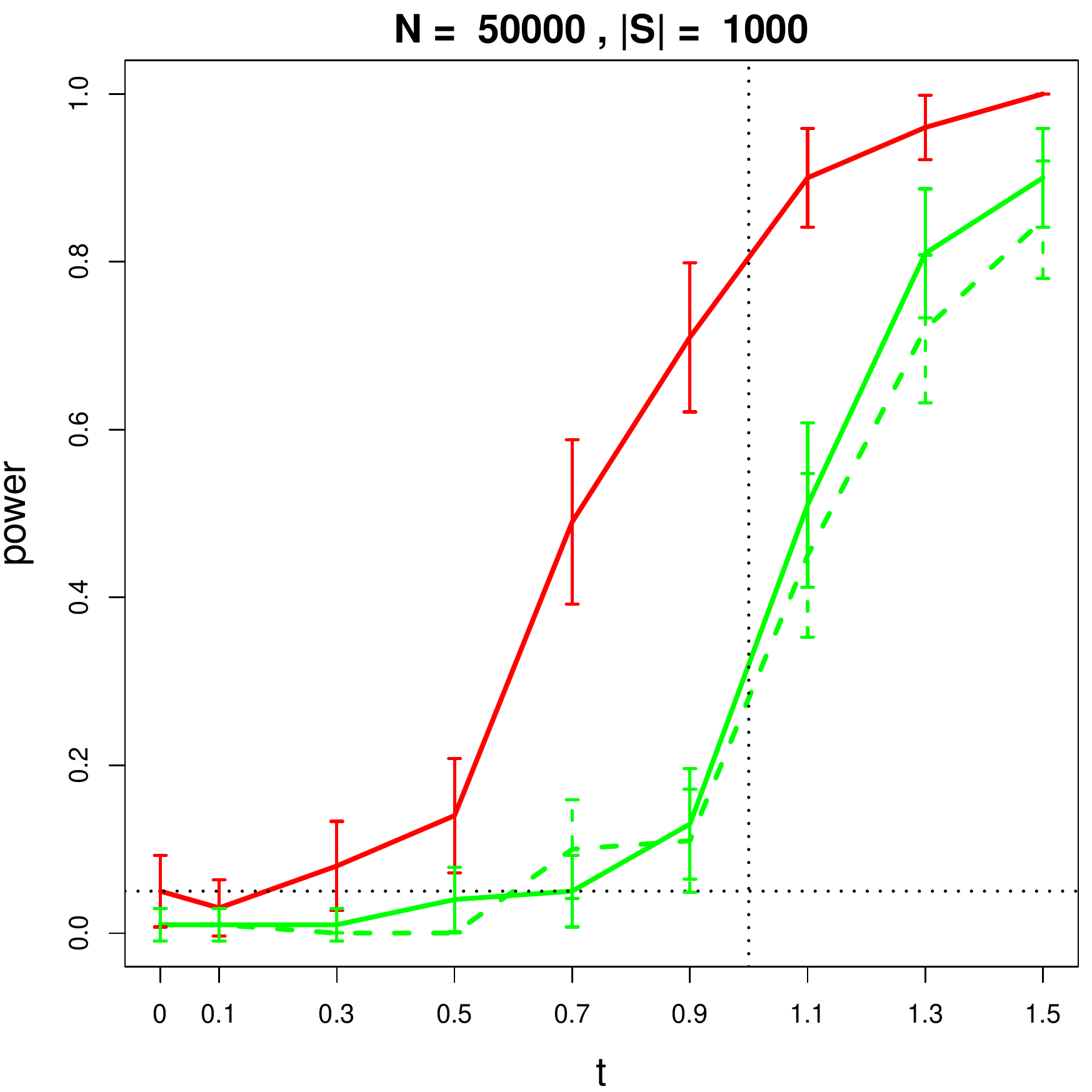}
\caption{Power curves (with 95\% margin of error) for the three tests as a function of the parameter $t$ in \eqref{t}: the rank scan test (red); RSI with bin size 10 (solid green); and RSI with bin size 20 (dashed green). The rank scan test is set at level 0.05 and its critical value is from 1000 repeats. On the left are the results for $|\cS| = 100$ and on the right for $|\cS| = 1000$.  $N = 50,000$ in both cases.  Each situation was repeated 100 times. The vertical black dashed line is the minimax threshold for $t$. The horizontal black dashed line is the significance level $0.05$.}\label{fig:detect}
\end{figure}

\paragraph{Detection}
In the problem of detection, we compare the performance of the rank scan test and RSI with bin size $m \in \{10, 20\}$ in normal data. To turn RSI into a test, we reject if it detects any anomalous interval.
In the simulation, we set sample size $N=50,000$ and considered the case where there is only one signal interval with known length $|\cS| \in \{100, 1000\}$.  The amplitude satisfy \eqref{t} as before.  We report the empirical power curves (based on 100 repeats) in \figref{detect}.

\vspace{1in}

To be fair, both methods only scan candidate signal intervals of length $|\cS|$. The rank scan is calibrated as before.
For RSI, we set the threshold to $\sqrt{2\log N}$ for the normalized data after localization to better control the family-wise type I error as  explained in \citep{tony2012robust}. 
From \figref{detect}, we can see that RSI is a bit more conservative.  In fact, a drawback of RSI is the difficulty to calibrate it correctly.\footnote{Of course, it could be calibrated by permutation, but this would make the procedure much more like the permutation scan test (with the same high-computational burden), somewhat far from the intentions of \citep{tony2012robust}.}  In any case, the rank scan test outperforms RSI in these simulations.

\paragraph{Identification}
In the problem of identification, we compare the rank scan and RSI. 
Although we focused on the problem of detection so far, a scan can be as easily used for testing as for estimation (i.e., identification).  Indeed, one sets an identification threshold and extract all the intervals that exceed that threshold.  Some post-processing --- such as merging significant intervals that intersect or keeping the most significant among significant intervals that intersect --- is often applied.   

Here, in an effort to be fair, we simply took the procedure of \citep{tony2012robust} --- which is essentially the procedure of \citep{jeng2010optimal} --- but calibrating as we did for testing.  Note that this implies a very stringent false identification rate (at the 0.05 testing level this means that the chances that one or more intervals are identified by mistake is 0.05).  We then compare its performance to that of the rank scan testing procedure calibrated in the same fashion.

Following \citep{tony2012robust}, in the simulation, we set the sample size to $N = 10^4$.  We consider a range of null distributions: the standard normal distribution, the $t$-distribution with 15 degrees of freedom and that with one degree of freedom. 
In each case, we set the signal mean to $\theta_{\cS} \in \{1, 1.5, 2\}$. There are three signal intervals, $\cS_1, \cS_2, \cS_3$, starting at positions 1000, 2000, 3000, and having lengths $2^4, 2^5, 2^6$, respectively. 
We set the threshold for the rank scan test by simulation at a significance level of $0.05$.  For RSI, we tried several bin sizes, $m \in \{2^3, 2^5\}$. 
To simplify the computation, both methods only scan dyadic intervals of length at most $2^6$. 
As in \citep{tony2012robust}, we compare their performance in terms of the following dissimilarities
\[
D_j = \min_{\hat{\cS} \in \hat{\mathbb{S}}} \{1 - \rho(S_j, \hat{\cS})  \},
\]
and the number of false positives, namely
\[
O = \{ \hat{\cS} \in \hat{\mathbb{S}}:  \hat{\cS} \cap \cS = \emptyset, \forall \cS \in \mathbb{S} \},
\]
where $\hat{\mathbb{S}}$ are the estimated signal intervals.

We report the average and standard deviation (in the parenthesis in the tables below) based on 200 repeats in Tables~\ref{tab:norm},~\ref{tab:t15}, and~\ref{tab:t1}.
We can see that the rank scan method performs better than RSI in when the null distribution is normal and $t(15)$, and it performs similarly to RSI with bin size $m=2^3$ in $t(1)$.  However, when the bin size of RSI is not properly chosen, RSI can perform poorly.

\begin{table}[h!]
\caption{Dissimilarity and number of over-selected intervals in $\cN(0,1)$} \label{tab:norm}
\bcenter
{\small 
\begin{tabular}{|cccccc|}
\hline
$\theta_{\cS}$ & Method & $D_1(|\cS_1|=2^4)$ & $D_2(|\cS_2|=2^5)$ & $D_3(|\cS_3|=2^6)$ & \#$O$ \\ \hline
&&&&& \\
  1  & Rank Scan& $0.734 ~(0.421)$ & $0.148 ~ (0.284)$ & $0.031 ~(0.049)$ & $0.000~ (0.000)$ \\
    & RSI($m=2^3$)& $0.916 ~(0.235)$ & $0.420~ (0.406)$ & $0.095~ (0.091)$ & $0.065~ (0.267)$ \\
    & RSI($m=2^5$)& $0.998 ~(0.029)$ & $0.959 ~(0.144)$ & $0.326 ~(0.278)$ & $0.130 ~(0.337)$ \\
    &&&&& \\
  1.5  & Rank Scan& $0.167 ~(0.326)$ & $0.019 ~(0.044)$ & $0.008 ~(0.012)$ & $0.000 ~(0.000)$ \\
    & RSI($m=2^3$)& $0.593 ~(0.391)$ & $0.132 ~(0.033)$ & $0.069 ~(0.029)$ & $0.080 ~(0.272)$ \\
    & RSI($m=2^5$)& $0.980 ~(0.087)$ & $0.729 ~(0.284)$ & $0.204 ~(0.044)$ & $0.025 ~(0.157)$ \\
    &&&&& \\
  2  & Rank Scan& $0.018 ~(0.051)$ & $0.006 ~(0.024)$ & $0.004 ~(0.008)$ & $0.000 ~(0.000)$ \\
    & RSI($m=2^3$)& $0.277~ (0.226)$ & $0.128~ (0.021)$ & $0.064 ~(0.013)$ & $0.065 ~(0.247)$ \\
    & RSI($m=2^5$)& $0.960 ~(0.122)$ & $0.476 ~(0.162)$ & $0.193 ~(0.032)$ & $0.010 ~(0.100)$ \\
&&&&& \\  \hline
\end{tabular}
}
\ecenter
\end{table}

\begin{table}[h!]
\caption{Dissimilarity and number of over-selected intervals in $t(15)$} \label{tab:t15}
\bcenter
{\small
\begin{tabular}{|cccccc|}
\hline
$\theta_{\cS}$ & Method & $D_1(|\cS_1|=2^4)$ & $D_2(|\cS_2|=2^5)$ & $D_3(|\cS_3|=2^6)$ & \#$O$ \\ \hline
&&&&& \\
1  & Rank Scan& $0.806 ~(0.369)$ & $0.223 ~(0.354)$ & $0.029~ (0.048)$ & $0.000 ~(0.000)$ \\
    & RSI($m=2^3$)& $0.926 ~(0.223)$ & $0.436~ (0.406)$ & $0.106~ (0.099)$ & $0.050~ (0.218)$ \\
    & RSI($m=2^5$)& $0.996 ~(0.041)$ & $0.944 ~(0.168)$ & $0.336 ~(0.278)$ & $0.125 ~(0.332)$ \\
 &&&&& \\
  1.5  & Rank Scan& $0.232 ~(0.378)$ & $0.026 ~(0.079)$ & $0.010 ~(0.017)$ & $0.000 ~(0.000)$ \\
    & RSI($m=2^3$)& $0.554 ~(0.391)$ & $0.143 ~(0.112)$ & $0.069 ~(0.031)$ & $0.075 ~(0.282)$ \\
    & RSI($m=2^5$)& $0.992 ~(0.057)$ & $0.732 ~(0.286)$ & $0.199 ~(0.042)$ & $0.020 ~(0.140)$ \\
&&&&& \\
  2  & Rank Scan & $0.034 ~(0.097)$ & $0.009 ~(0.019)$ & $0.005 ~(0.014)$ &  $0.000 ~(0.000)$ \\
    & RSI($m=2^3$)& $0.277~ (0.220)$ & $0.128~ (0.022)$ & $0.063~ (0.013)$ & $0.060 ~(0.238)$ \\
    & RSI($m=2^5$)& $0.968 ~(0.107)$ & $0.521~(0.214)$ & $0.192 ~(0.030)$ & $0.010 ~(0.100)$ \\ 
&&&&& \\  \hline
\end{tabular}
}
\ecenter
\end{table}

\begin{table}[h!]
\caption{Dissimilarity and number of over-selected intervals in $t(1)$} \label{tab:t1}
\bcenter
{\small 
\begin{tabular}{|cccccc|}
\hline
$\theta_{\cS}$ & Method & $D_1(|\cS_1|=2^4)$ & $D_2(|\cS_2|=2^5)$ & $D_3(|\cS_3|=2^6)$ & \#$O$ \\ \hline
&&&&& \\
 1  & Rank Scan& $0.989 ~ (0.082)$ & $0.878~ (0.305)$ & $0.461~ (0.448)$ & $0.000~ (0.000)$ \\
    & RSI($m=2^3$)& $0.950~ (0.186)$ & $0.764~ (0.370)$ & $0.332~ (0.358)$ & $4.305 ~(5.653)$ \\
    & RSI($m=2^5$)& $0.998 ~(0.022)$ & $0.982 ~(0.098)$ & $0.609 ~(0.392)$ & $ 0.520 ~(0.501)$ \\
  &&&&& \\
  1.5  & Rank Scan& $0.922 ~(0.251)$ & $0.542 ~(0.455)$ & $0.067 ~(0.132)$ & $0.000 ~(0.000)$ \\
    & RSI($m=2^3$)& $0.843 ~(0.307)$ & $0.342 ~(0.354)$ & $0.104 ~(0.080)$ & $3.920 ~(2.082)$ \\
    & RSI($m=2^5$)& $0.983 ~(0.079)$ & $0.877 ~(0.236)$ & $0.225 ~(0.111)$ & $0.055 ~(0.229)$ \\
&&&&& \\
  2  & Rank Scan& $0.763 ~(0.410)$ & $0.206 ~(0.333)$ & $0.043 ~(0.093)$ & $0.000 ~(0.000)$ \\
    & RSI($m=2^3$)& $0.619~ (0.382)$ & $0.154~ (0.121)$ & $0.089~ (0.063)$ & $3.945~ (2.385)$ \\
    & RSI($m=2^5$)& $0.978 ~(0.090)$ & $0.667 ~(0.280)$ & $0.208 ~(0.05)$ & $0.060 ~(0.238)$ \\ &&&&& \\  \hline
\end{tabular}
}
\ecenter
\end{table}

\subsection{Application to the real data} \label{sec:CNV}
In this section, we apply the methods to the problem of detecting the copy number variant (CNV) in the context of next generation sequencing data. 
We compare the rank scan method and RSI on the task of identifying  short reads on chromosome 19 of a HapMap Yoruban female sample (NA19240) from the 1000 genomes project (\url{http://www.1000genomes.org}), which is the same data set used in \citep{tony2012robust}. 
Following standard protocols \citep{ernst2011mapping}, we extend all the reads to 100 base pairs (BPs).  We take $10^6$ reads from the whole data set for comparison purposes resulting in 1,281,502 genomic locations.

We tune RSI as done in \citep{tony2012robust}, setting the bin size to $m = 400$ and the maximum BPs in a possible CNV to $L = 2^{16}$.  Note that \citep{tony2012robust} took $L = 60,000$, which is a bit smaller than $2^{16}$.  (We chose the latter because we only scan intervals of dyadic length.) 
To save computational time, in the implementation of the rank scan  we group read depths in every 200 positions and take the summation of the read depths for each bin and use that as input (meaning, we rank the sums and scan the ranks). 
%However, one can definitely apply the rank scan method to the original scale with a proper scanning range. 
We get the critical value for the rank scan method under the significance level $0.05$ from 1000 repeats. In the experiment, we let RSI and the rank scan method only scan dyadic intervals of lengths from $2^1$ to $2^{16}$.

\begin{figure}[h!]
\centering
\includegraphics[scale=0.6]{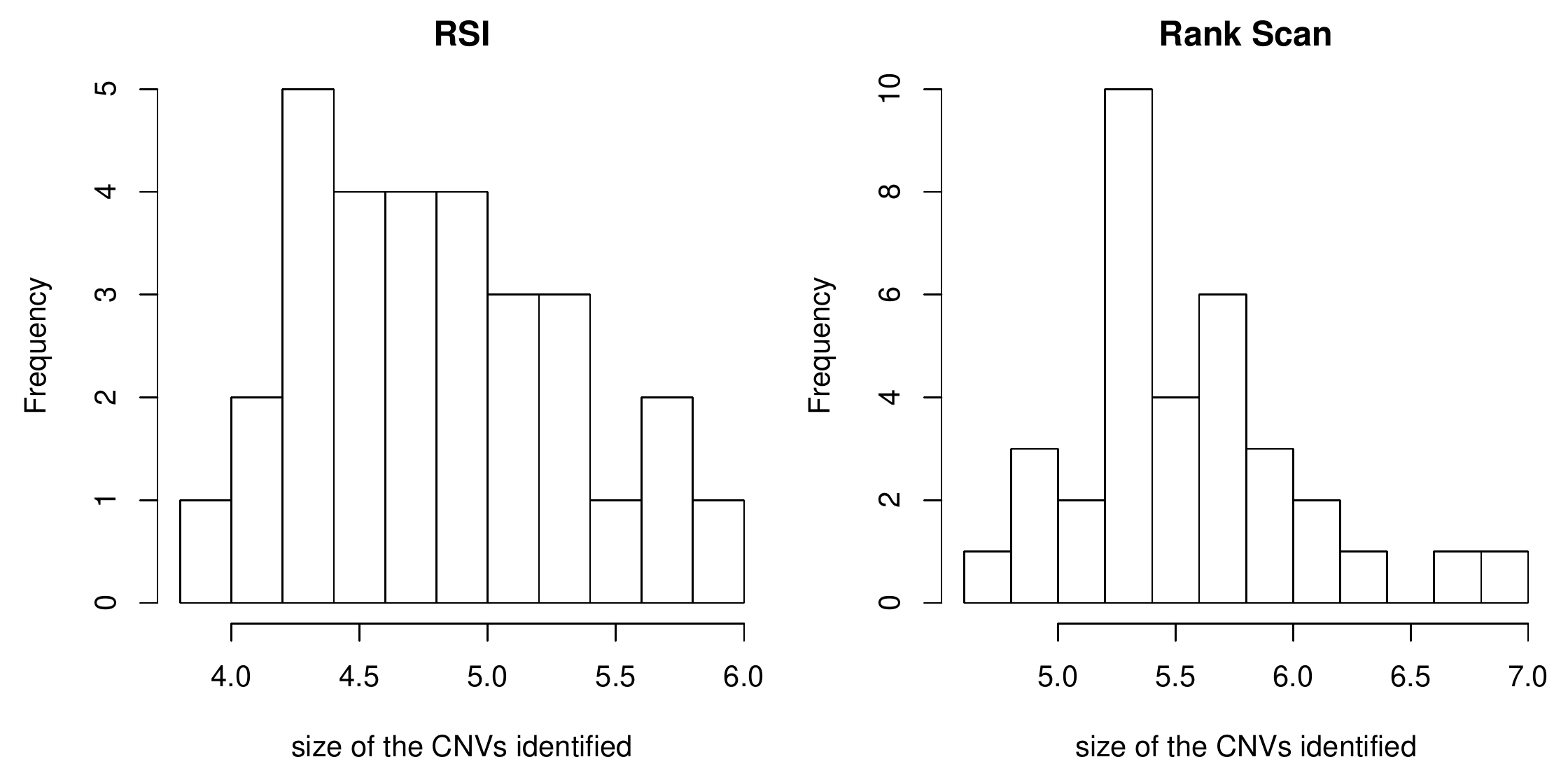} 
\caption{Histogram of the read depths of the selected CNVs in log scale (base 10). Both methods only scan dyadic intervals of lengths from $2^1$ to $2^{16}$.  The RSI used a bin size $m = 400$, while the rank scan was calibrated as for testing.}
\label{fig:hist_compare}
\end{figure}

After merging the contiguous selected segments, RSI found 30 possible CNVs and the rank scan method selected 34. \figref{hist_compare} shows the histograms of the read depths of the selected CNVs. We can see the read depth in the rank scan method is generally larger than that in RSI. 
%A close comparison on positions from the $5 \cdot 10^5$ to $7 \cdot 10^5$ is indicated in \figref{readdepth_compare}.  
%In the end, it is difficult to conclude which is method comes out best without confirmation from biological experiments.  

\section{Discussion} \label{sec:discussion}

In this paper we consider a prototypical structured detection setting with the particularity that the null distribution is unknown.  
When the null distribution is known, various works have shown that a form of scan test achieves the best possible asymptotic power. 
When the null distribution is unknown, one can alternatively calibrate the scan test by permutation.  This has been suggested a number of times in the detection literature. 
\thmref{perm_test} implies doing this results in no loss of asymptotic power compared to a calibration by Monte Carlo with full knowledge of the null distribution. 
To circumvent the expense of calibrating by permutation, we propose to scan the ranks. 
\thmref{rank_test} and \prpref{rank_test_exp} imply that this results in very little loss in asymptotic power. 
In our empirical experiments all three methods perform comparably. 
Generalizations to multivariate scenarios are also possible (e.g., $X_v\in\bbR^d$ with $d>1$). The exact procedure will depend heavily on the specific problem context.  For instance, in imaging contexts the entries of $X_v$ correspond to measurements in different wavelengths that might be suitably combined in a single univariate score.

\medskip\noindent {\em Censoring before permutation.}
When $F_0$ is not of compact support, we can enforce it by applying a censoring of the form $\tilde X_v = X_v \IND{|X_v| \le t} + t \sign(X_v) \IND{|X_v| > t}$.  With a choice of threshold $t = t_N \to \infty$ slowly (e.g., $t_N = \log\log N$), \thmref{perm_test} applies with $\frac{1}{|\cS|}\sum_{v\in\cS} \theta_v$ replaced by $\min_{v\in\cS} \theta_v$ and without an upper bound on the $\theta_v's$. The proof of this result is nearly identical except for very minor modifications.
This censoring has the added advantage of making the method more robust to possible outliers.

\medskip\noindent {\em Other scoring functions.}
Although rank-sums are intuitive and classically used, any scan based on $h(r_v)$, where $h$ is increasing, is valid.  (Recall that $r_v$ is the rank of $x_v$ in the sample.)  In two-sample testing, it is known that there is no uniformly better choice of function $h$.  See \cite[Sec 6.9]{MR2135927} where it is shown that choosing $h(r) = \E(Z_{(r)})$ --- where $Z_{(1)} < \cdots < Z_{(N)}$ are the order statistics of a standard normal sample --- is (in some sense) optimal in the normal location model.
Our method of proof applies to a general $h$.

\medskip\noindent {\em Unstructured subsets.}
No permutation approach (including a rank-based approach) has any power for detecting unstructured anomalies.  A prototypical example is when $\bbS$ is the class of all subsets, or all subsets of given size, the latter including the class of singletons.

\section{Proofs} \label{sec:proofs}

\subsection{Proof of \thmref{perm_test}}

Suppose first we are under the null hypothesis. Note that $\bX=(X_v, v \in \cV)$ are IID under the null, and therefore exchangeable. This means that, for any permutation $\pi$ the marginal distributions of $\scan(\bX)$ and $\scan(\bX_\pi)$ are the same. This implies that $\scan(\bX)$ is conditionally uniformly distributed on the set $\{ \scan(\bX_\pi), \pi \in \cV ! \}$ (with multiplicities) and so
\[
\P \left(\left| \{\pi \in \cV! : \scan(\bX_\pi) \ge \scan(\bX)\} \right| \leq \alpha \cV !\right) \leq \frac{\lfloor \alpha |\cV|! \rfloor}{|\cV|!} \le \alpha\ ,
\]
where $\lfloor z \rfloor$ denotes the integer part of $z$. If there were no ties, the first inequality above would be an equality, but with ties present the test becomes more conservative. For more details on permutation tests the reader is referred to \citep{MR2135927}.

All that remains to be done is to study the permutation test under the alternative hypothesis. This requires two main steps. First we need to control the randomness in the permutation, conditionally on the observations $\bx$. Once this is done we remove the conditioning. 

The key to the first step is the following Bernstein's inequality for sums of variables sampled without replacement from a finite population.

\begin{lem}[Bernstein's inequality for sampling without replacement] \label{lem:bernstein}
Let $(Z_1, \dots, Z_m)$ be obtained by sampling without replacement from a given a set of real numbers $\{z_1, \dots, z_J\} \subset \bbR$.  Define 
$z_{\rm max} = \max_j z_j$, $\bar z = \frac1J \sum_j z_j$, and $\sigma_z^2 = \frac1J \sum_j (z_j - \bar z)^2$.
Then the sample mean $\bar Z = \frac1m \sum_i Z_i$ satisfies 
\[
\P\big(\bar Z \ge \bar z + t\big) 
\le \exp\Bigg[- \frac{m t^2}{2 \sigma_z^2 + \frac23 (z_{\rm max} - \bar z) t}\Bigg], \quad \forall t \ge 0.
\]
\end{lem}

This result is a consequence of \citep[Th.~4]{hoeffding} and Chernoff's bound, from which Bernstein's inequality is derived, as in\footnote{There is a typo in the statement of the result in \cite[p 851]{ShoWel}, but following the proof one can find the correct result. Where the statement of the result reads $- \frac{\lambda}{2 \sigma^2}$ we should have $-\frac{\lambda^2}{2 \sigma^2}$ instead} \cite[p 851]{ShoWel}. 
See \citep{boucheron2013concentration,bardenet2013concentration,MR2571413} for a discussion of the literature on concentration inequalities for sums of random variables sampled without replacement from a finite set.

Applying this result for a fixed (but arbitrary) set $\cS^* \in \bbS_b$ when $\pi$ is uniformly drawn from $\cV!$ and $\bx$ is given, we get
\[
\P \left( Y_{\cS^*}(\bx_\pi )-\sqrt{|\cS^*|}\bar x \ge t \right) 
\le \exp\Bigg[- \frac{t^2}{2 \sigma_x^2 + \frac23 (x_{\rm max} - \bar x) t/\sqrt{|\cS^*|}}\Bigg], \quad \forall t \ge 0,
\]
using the same notation as in \lemref{bernstein}.  
Plugging in $t=\scan(\bx)$, noting that $|\cS^*| \geq 2^{q_l}/(1+2^{-b+2}) \ge 2^{q_l}/2$ eventually (because $b \to \infty$), and using this together with a union bound, we get
\beq \label{P-bound}
\mathfrak{P}(\bx) \le |\bbS_b | \exp\Bigg[- \frac{\scan(\bx)^2}{2 \sigma_x^2 + (x_{\rm max} - \bar x) 2^{-q_l/2} \scan(\bx)}\Bigg].
\eeq
(The $\frac23$ in the denominator, when multiplied by $\sqrt{2}$, from $|\cS| \ge 2^{q_l}/2$, is still less than 1.)

Now we proceed by upper bounding the right-hand side of the above inequality by assuming we are under the alternative, which yields an upper bound for the P-value $\mathfrak{P}(\bX)$. This amounts to controlling the terms $X_{\rm max} - \bar X$, $\sigma_{\bX}^2$ and $\scan(\bX)$ under the alternative (upper-case $X$ relates to the random quantities.)

Recall that $F_0$ has zero mean and unit variance and note that $\E_\theta (X)$ and $\Var_\theta (X)$ are continuous in $\theta$ (and thus bounded on the interval $[0,\tilde{\theta}]$).

%%%%%%%%%%%%%%%%

We begin by controlling $\bX_{\rm max} - \bar{\bX}$.  Let $S$ denote the anomalous interval under the alternative.  We have 
\[
\bar{\bX} = \frac{1}{N} \sum_{v\in\cV} \E (X_v) \ +\  \frac{1}{N} \sum_{v\in\cV} (X_v - \E (X_v)) = O(|\cS|/N) + o_P(1) = o_P(1)\ ,
\] 
as $N\to \infty$, since $|\cS| = o(N), \theta_v \leq \tilde{\theta}$ for all $v\in \cV$, and using Chebyshev's inequality in the second equality.  Furthermore, let $\bX_{\rm max, S}=\max_{v\in \cS} X_v$ be the maximum over the anomalous set $\cS$.  Let $\bar \cS$ denote the complement of $\cS$.  A union bound together with $\bX_{\rm max} = \bX_{\rm max, \cS} \vee \bX_{\rm max, \bar{\cS}}$ implies
\[
\P (\bX_{\rm max} > x) \leq \P (\bX_{\rm max,\cS} > x) + \P (\bX_{\rm max,\bar{\cS}} > x) \leq |\cS| \bar{F}_{\tilde{\theta}}(x) + |\bar{\cS}| \bar{F}_0(x) \ ,
\]
where $\bar{F}_\theta(x) = \P_\theta (X > x)$ and we used the fact that $\bar{F}_\theta(x)$ is monotone increasing in $\theta$ - see \secref{exponential}.  For $c \in (0, \theta_\star - \tilde{\theta})$, we have
\begin{align*}
\bar F_{\tilde\theta}(x) & = \int_x^\infty e^{\tilde\theta u - \log \varphi_0(\tilde\theta)} {\rm d} F_0 (u) \\
& = \frac{1}{\varphi_0(\tilde\theta)} \int_x^\infty e^{-c u} e^{(\tilde\theta +c) u} {\rm d} F_0 (u) \leq \frac{\varphi_0 (\tilde\theta +c)}{\varphi_0 (\tilde\theta)} e^{-c x} \ .
\end{align*}
Using this with the above union bound gives $\P(\bX_{\rm max} > (2/c) \log N) \to 0$ as $N\to \infty$.  This and the bound on $\bar{\bX}$ imply that
\[
\P(\bX_{\rm max} -\bar{\bX} > (3/c) \log N) \to 0\ .
\]

We now consider $\sigma_\bX^2$. Similarly as before, we have
\begin{align*}
\sigma_\bX^2 = \frac{1}{N} \sum_{v\in\cV} (X_v - \bar X)^2 \leq \frac{1}{N} \sum_{v\in\cV} X_v^2
& = \frac{1}{N} \sum_{v\in\cV} \E (X_v^2) + \frac{1}{N} \sum_{v\in\cV} (X_v^2 - \E (X_v^2)) \ .
\end{align*}
On one hand,
\begin{align*}
\frac{1}{N} \sum_{v\in\cV} \E (X_v^2) & = \frac{1}{N} \sum_{v\notin \cS} \Var (X_v) + \frac{1}{N} \sum_{v \in \cS} (\Var (X_v) + \E (X_v)^2) \\
& = 1 - \frac{|\cS |}{N} + O\left( \frac{|\cS|}{N} \right) = 1 + o(1)\ ,
\end{align*}
using $\Var (X_v) = 1$ for $v\notin \cS$, $\max_{v\in \cS} \Var (X_v) < \infty$ and $\max_{v\in \cS} \E (X_v) < \infty$ (since $\max_{v\in \cS} \theta_v \leq \tilde{\theta}$), as well as our assumption that $|\cS| = o(N)$.
On the other hand,
\[
\frac{1}{N} \sum_{v\in\cS} (X_v^2 - \E (X_v^2)) = O_P(1/\sqrt{N}) \ ,
\]
using the fact that $\max_{v\in\cS} \E (X_v^4) < \infty$ (since $\max_{v\in\cS} \theta_v \leq \tilde{\theta}$) combined with Chebyshev's inequality. We may therefore conclude that
\[
\P (\sigma_\bX^2 \leq 1+\eps /4 ) \to 1\ ,
\]
with a fixed but arbitrary $\eps>0$ (we will choose an appropriate value for $\eps$ later on).

From Lemma~\ref{lem:approximating_net} (which does apply to the newly defined $\bbS_b$) there is a set $\cS^* \in \bbS_b$ such that $\cS \subseteq \cS^*$ and $\rho(\cS,\cS^*)\geq (1 + 2^{-b+2})^{-1/2}$.  Note that $\rho(S,S^*) = 1 - o(1)$ by the fact that $b \to \infty$. We then have
\begin{align*}
\scan (\bX ) & \geq \bX_{\cS^*}-\sqrt{|\cS^* |}\bar{\bX} = \sqrt{|\cS^* |} (\bar{\bX}_{\cS^*} - \bar{\bX}) \\
& \geq \sqrt{|\cS^* |} \left( \frac{|\cS |(N-|\cS^* |)}{|\cS^* |N} \bar{\bX}_\cS - \frac{N-|\cS |}{N} \bar{\bX}_{\cV \setminus \cS} \right) \ ,
\end{align*}
where $\bar{\bX}_\cS$ and $\bar{\bX}_{\cV \setminus \cS}$ are the averages of the components of $\bX$ over the sets $\cS$ and $\cV \setminus \cS$ respectively. By Chebyshev's inequality, 
\begin{align*}
\bar{\bX}_\cS & = \frac{1}{|\cS |} \sum_{v \in \cS} \E (X_v) + O_P(1/\sqrt{|\cS |})\ , \\
\bar{\bX}_{\cV \setminus \cS} & = O_P(1/\sqrt{N-|\cS |}) \ .
\end{align*}

Recall that we have
\begin{equation}\label{Rank_theta_ddag}
\frac{1}{|\cS|} \sum_{v\in \cS} \theta_v \geq \tau \sqrt{\frac{2 \log N}{|\cS|}}:=\theta_\ddag~.
\end{equation}
Note that $\theta_\ddag$ converges to zero by the assumption on $q_l$ and the fact that $\tau$ is fixed.  Furthermore $\E_{\theta}(X)$ is increasing in $\theta$ (as $\tfrac{\partial}{\partial \theta} \E_{\theta}(X) = \E_{\theta}(X^2) \geq 0$) and $\E_{\theta}(X)=\theta + O(\theta^2)$ when $\theta \to 0$ (this can be checked by noting $\E_{\theta}(X) = \int xe^{\theta x} {\rm d}F_0 (x)$ and writing the Taylor expansion of $e^{\theta x}$ around zero). Thus $\tfrac{1}{|\cS |} \sum_{v\in \cS} \E (X_v) \geq \E_{\theta_\ddag} (X) = \theta_\ddag + O(\theta_\ddag^2)$ because $\theta_\ddag \to 0$.  Using $\sqrt{|\cS^*|}=(1+o(1))\sqrt{|\cS|}$ and $|\cS|=o(N)$ we get
\[
\scan (\bX ) \geq (1+o(1)) \tau \sqrt{2 \log N} + O_P(1)\ ,
\]
therefore
\[
\scan (\bX ) \geq \sqrt{2 (1+\eps /2) \log N}\ ,
\]
with probability tending to one as $N \to \infty$, where we take $\eps$ so that $\tau = \sqrt{1+\eps}$.

We are ready to make use of the upper bound on the P-value given by \eqref{P-bound} and using the condition on $q_l$ we get
\begin{align*}
\log \mathfrak{P}(\bX ) & \leq \log |\bbS_b| - \frac{2(1+\eps /2) \log N}{2 (1+\eps /4)  + (3/c)(\log N) \sqrt{2^{-q_l+1} (1+\eps /2) \log N}} \\
& \leq \log |\bbS_b | - \frac{(1+\eps /2) \log N}{1+\eps /4 + o(1)} \ ,
\end{align*}
with probability going to 1. For the size of the approximating net we have 
\begin{equation} \label{log-S}
\log |\bbS_b | \leq \log \left( N 4^{b+1} \right) = \log N + (b+1) \log 4 = (1 + o(1)) \log N\ ,
\end{equation}
by our assumption on $b$. Combining these allows us to conclude that $\log \mathfrak{P}(\bX)\to -\infty$ (meaning $\mathfrak{P}(\bX)\to 0$) with probability tending to one, implying that the test has power tending to 1 as $N \to \infty$, concluding the proof.

%%%%%%%%%%%%%%%%%%%%%%%%%%%%%%%%%%%%
%%%%%%%%%%%%%%%%%%%%%%%%%%%%%%%%%%%%
%%%%%%%%%%%%%%%%%%%%%%%%%%%%%%%%%%%%

\subsection{Proof of \thmref{rank_test}}

The arguments used for the general permutation test apply verbatim under the null hypothesis, so all that remains to be done is to study the performance of the rank scan test under the alternative.

We may directly apply \eqref{P-bound}, to obtain
\begin{equation}\label{eqn:R_Bernstein}
\mathfrak{P}(\br) \le |\bbS_b| \exp\Bigg(- \frac{\scan(\br)^2}{\frac{N^2}{6} + \frac{N}{2} 2^{-q_l/2} \scan(\br)}\Bigg),
\end{equation}
where we used $\sigma_r^2 = (N^2-1)/12 < N^2 /12$, $r_{\rm max} = N$ and $\bar r = (N+1)/2$, so that $r_{\rm max} - \bar{r} <N/2$. The previous bounds can be directly computed when there are no ties in the ranks, and it is easy to verify that they also hold if ties are dealt with in any of the classical ways (assigning the average rank, randomly breaking ties, etc).  As before, this is a result conditional on the observations $\bX =\bx$ and hence the ranks $\bR = \br$. The next step is to remove this conditioning, which now amounts to controlling the term $\scan (\bR )$.

Let $\cS$ denote the anomalous interval under the alternative. From Lemma~\ref{lem:approximating_net} there is a set $\cS^* \in \bbS_b$ such that $\cS \subseteq \cS^*$ and $\rho(\cS,\cS^*)\geq (1 + 2^{-b+2})^{-1/2}$, therefore $\rho(\cS,\cS^*) = 1 - o(1)$ by the fact that $b \to \infty$. Since 
\[\scan (\bR) \geq Y_{\cS^*}(\bR)-\sqrt{|\cS^*|}\tfrac{N+1}{2}\ ,\]
we focus on obtaining a lower bound on $Y_{\cS^*}(\bR)$ that applies with high probability.

Note that
\[\E (Y_{\cS^*}(\bR)) = \frac{1}{\sqrt{|\cS^*|}} \sum_{v\in \cS^*} \E (R_v)\ ,\]
and
\[\Var (\bR_{\cS^*}) = \frac{1}{|\cS^*|} \left( \sum_{v\in \cS^*} \Var (R_v) + \sum_{v,w\in \cS^*, v \neq w} \Cov (R_v,R_w) \right)\ .\]
In an analogous fashion to that in \citep{MR758442}, we can make the following claims about the first two moments of the ranks.

\begin{lem}\label{lem:rank-sum}
Suppose $Z_i \sim F_i, i\in [s]$ and independent, also independent of $\{ Z_i \}_{i\in [s+1,n]}$ which are i.i.d. and distributed as $F_0$. Let $R_i$ denote the rank (in increasing order) of $Z_i$ in the combined sample, and suppose ties are broken randomly. Define
\[
p_{i,j} = \P (X>Y) + \tfrac{1}{2} \P (X=Y) \ ,
\]
where $X\sim F_i, Y\sim F_j$ are independent. For $i\in [s]$
\[
\E (R_i) = \left\{
\begin{array}{ll}
(n-s) p_{i,0} + \displaystyle{\sum_{j\in [s],j \neq i}} p_{i,j} +1 & \textrm{, when $i\in [s]$,}\\
\frac{n+s+1}{2} - \displaystyle{\sum_{j\in [s]}} p_{j,0} & \textrm{, when $i\notin [s]$.}
\end{array} \right.
\]
Furthermore, as $n,s\to \infty, s=o(n)$, for $i \in [s]$
\[
\Var (R_i) = (\lambda_i - p_{i,0}^2) n^2 + O(sn) \ ,
\]
where
\[
\lambda_i = \P (\{ X > Y_1\} \cap \{ X > Y_2\} ) + \P (X = Y_1 > Y_2) + \tfrac{1}{3} \P (X = Y_1 = Y_2)\ ,
\]
where $X\sim F_i$ and $Y_1 ,Y_2 \sim F_0$ are jointly independent. Finally, for any $i,j \in [n]$
\[
\Cov (R_i ,R_j) = O(n) \ .
\]
\end{lem}

For the sake of completeness we sketch a proof of Lemma~\ref{lem:rank-sum} in Appendix~\ref{sec:proof-rank-sum}. Recall the definition of $p_v$ in \eqref{p_v} and $p_{v,w}$ in Lemma~\ref{lem:rank-sum}. 
Using the fact that for any $i,j$ we have $p_{i,j} + p_{j,i} =1$ we get
\begin{align*}
\sqrt{|\cS^* |} \E (Y_{\cS^*} (\bR )) 
& = \sum_{v\in \cS^*} \E (R_v)
= \sum_{v\in \cS} \E (R_v) + \sum_{v\in \cS^* \setminus \cS} \E (R_v)\\
& = \sum_{v\in \cS} \Big( (N-|\cS |) p_v + \sum_{w\in \cS ,w\neq v} p_{v,w} + 1 \Big) + \sum_{v\in \cS^* \setminus \cS} \Big( \tfrac12 (N+|\cS |+1) + \sum_{w\in \cS} p_w \Big) \\
& = |\cS | (N-|\cS |) \bar{p}_\cS + \sum_{v\in \cS } \sum_{w\in \cS ,\, w\neq v} p_{v,w} +|\cS | + |\cS^* \setminus \cS | \tfrac12 (N+|\cS |+1) - |\cS^* \setminus \cS | |\cS |\bar{p}_\cS \\
& = |\cS | (N-|\cS |-|\cS^* \setminus \cS |) \bar{p}_\cS + \tfrac12 |\cS |(|\cS |+|\cS^* \setminus \cS |) + \tfrac12 |\cS | + |\cS^* \setminus \cS | \tfrac{N+1}{2} \\
& = |\cS | (N-|\cS |-|\cS^* \setminus \cS |) (\bar{p}_\cS -1/2) + |\cS |\tfrac{N+1}{2} + |\cS^* \setminus \cS | \tfrac{N+1}{2} \\
& = |\cS | (N-|\cS |-|\cS^* \setminus \cS |) (\bar{p}_\cS -1/2) + |\cS^* | \tfrac{N+1}{2} \ ,
\end{align*}
where $\bar{p}_\cS = \tfrac{1}{|\cS |} \sum_{v\in \cS} p_v$ is the average of $p_v$ over the anomalous set.

Note that for any $v\in [N]$ we trivially have $\Var (R_v) \leq N^2$, and by Lemma~\ref{lem:rank-sum}, $\Cov (R_v ,R_w) = O(N)$, so $\Var (Y_{\cS^*} (\bR )) = O(N^2)$. Hence, using Chebyshev's inequality we obtain
\begin{align}
Y_{\cS^*} (\bR ) -\sqrt{|\cS^* |}\tfrac{N+1}{2} & = \frac{|\cS |}{\sqrt{|\cS^*|}} (N-|\cS |-|\cS^* \setminus \cS |) (\bar{p}_\cS -1/2) + O_P(N)  \label{eqn:bound_Y}\\
& \geq \rho (\cS ,\cS^* ) (N-o(N)) \tau \sqrt{2 \log N} + O_P(N) \ ,\nonumber
\end{align}
where we used the condition on $q_u$ to conclude that $|\cS^*|+|\cS^* \setminus \cS |=o(N)$. In summary we have
\[
\scan (\bR ) \geq  c \frac{N}{2\sqrt{3}} \sqrt{2 \log N} \ ,
\]
with probability going to 1 as $N\to \infty$, where $c\in (1,2\tau \sqrt{3})$.

Plugging this back into \eqref{eqn:R_Bernstein} and accounting for the condition on $q_l$ we get
\begin{align*}
\log \mathfrak{P}(\bR ) & \leq \log |\bbS_b| - \frac{c^2\frac{N^2}{6} \log N}{\frac{N^2}{6} + \frac{N^2}{2} \frac{c}{2\sqrt{3}} \sqrt{2^{-q_l+1} \log N}} \\
& \leq \log |\bbS_b| - \frac{c^2 \log N}{1+o(1)} \ ,
\end{align*}
with probability going to 1. Noting that the upper bound on $|\bbS_b |$ in \eqref{log-S} still holds and that $c>1$ allows us to conclude that $\log \mathfrak{P}(\bR ) \to -\infty$ as $N\to \infty$, hence the test is asymptotically powerful.

\subsection{Proof of \prpref{rank_test_exp}}

Showing this result amounts to relate $p_v\equiv p_{\theta_v}$ with $\theta_v$. This is conveniently done by a Taylor expansion around zero. For ease of presentation let $\theta\equiv\theta_v$ in what follows.  When $F_0$ is discrete, we have
\[
p_\theta = \int \left( \bar{F}_\theta (x) +\tfrac{1}{2} f_\theta (x) F_0 (x) \right) {\rm d} F_0(x)\ .
\]
We expand the integrand seen as a function of $\theta$ around $\theta = 0$ up to a second order error term. 
We have 
\[
\tfrac{\partial}{\partial \theta} f_\theta (x) \Bigg|_{\theta=0} = x, \quad
\tfrac{\partial}{\partial \theta} \bar F_\theta (x) \Bigg|_{\theta=0} = \int_{(x,\infty )} u\ {\rm d} F_0 (u)\ ,
\]
where the second identity comes from differentiating inside the integral defining $\bar{F}_\theta$, justified by dominated convergence. Note that $\tfrac{\partial^2}{\partial \theta^2} f_\theta (x)$ is integrable w.r.t.~$F_0$ when $\theta \in [0,\theta^*)$ and the same holds for $\tfrac{\partial^2}{\partial \theta^2} \bar{F}_\theta (x)$ as well. Hence let 
\begin{align*}
c'_0 & := \int \sup_{\tilde{\theta} \in [0,\theta]} \tfrac{\partial^2}{\partial \theta^2} f_\theta (x) \Bigg|_{\theta=\tilde{\theta}} {\rm d} F_0 (x) < \infty \ , \textrm{ and}\\ 
c_0 & := \int \sup_{\tilde{\theta} \in [0,\theta]} \tfrac{\partial^2}{\partial \theta^2} \bar F_\theta (x) \Bigg|_{\theta=\tilde{\theta}} {\rm d} F_0 (x) < \infty \ .
\end{align*}
Therefore 
\begin{align*}
p_\theta & \geq \int \bar{F}_0(x)+\tfrac{1}{2}F_0 (x) + \theta \left( \int_{(x,\infty )} u\ {\rm d} F_0 (u)  + \tfrac{1}{2} F_0 (x) x \right) {\rm d} F_0 (x) - \tfrac{\theta^2}{2} (c_0 + c'_0 /2) \\
& = p_0 + \theta \big(\E_0 ( X \IND{X>Y} ) + \tfrac{1}{2} \E_0 ( X \IND{X=Y} ) \big) - \tfrac{\theta^2}{2} (c_0 +c'_0 /2) \\
& = \tfrac{1}{2} + \theta \Upsilon_0 - \tfrac{\theta^2}{2} (c_0 +c'_0 /2)\ .
\end{align*}
When $F_0$ is continuous, we have
\[p_\theta = \int \bar{F}_\theta (x) {\rm d} F_0 (x)\ ,\]
and similar calculations lead to
\[p_\theta \geq \tfrac{1}{2} + \theta \Upsilon_0 - \tfrac{\theta^2}{2} c_0 \ .\]

In summary, we conclude that $p_\theta \geq \tfrac{1}{2} + \theta \Upsilon_0 + O(\theta^2)$ as $\theta\to 0$. In addition, note that $p_\theta$ is monotonically increasing in $\theta$, by virtue of the fact that $(F_\theta: \theta \ge 0)$ has monotone likelihood ratio.  Therefore,
% , for any $v\in\cS$ we have
% $$p_{\theta_v} \geq \tfrac{1}{2} + \Upsilon_0 \tau_0\sqrt{\frac{2\log N}{|\cS|}} + O\left(\frac{2\log N}{|\cS|}\right)\ .$$
\[
\frac{1}{|\cS|}\sum_{v\in\cS} p_{\theta_v} \geq \frac{1}{2} + \tau\Upsilon_0 \sqrt{\frac{2 \log N}{|\cS |}} + O\left(\frac{2 \log N}{|\cS |}\right) \ .
\]

Finally, using the above bound in \eqref{eqn:bound_Y} and proceeding in an analogous fashion as in Theorem~\ref{thm:rank_test} yields the desired result.

\subsection{Proof of \prpref{small}}

We treat each case separately.

\medskip\noindent {\em Condition (i).} 
The same arguments hold as before under the null, so again we are left with studying the alternative. To deal with smaller intervals, we need a slightly different concentration inequality than before.

\begin{lem}[Chernoff's inequality for ranks] \label{lem:chernoff}
In the context of Lemma~\ref{lem:bernstein}, assume that $z_j = j$ for all $j$.  Then
\[
\P \left( \bar Z \ge \bar z + t \right) \leq \exp \left( - m\, {\textstyle\sup_{\lambda \ge 0}} \psi(t,\lambda )\right) \ , \quad \forall t \ge 0\ ,
\]
where
\[
\psi(t,\lambda ) := \lambda t - \log \left( \frac{\sinh (\lambda n/2)}{n \sinh (\lambda /2)} \right) \ . 
\]
\end{lem}

Similarly to Lemma~\ref{lem:bernstein} this result is also a consequence of Theorem~4 of \cite{hoeffding} and Chernoff's bound. However, with the assumption on $z_j$ in the lemma above we can directly compute the moment generating function of $Z_j$ after using Chernoff's bound instead of upper bounding it, as is classically done to obtain Bernstein's inequality.

In the present context, this yields
\[
\mathfrak{P}(\br ) \leq |\bbS | \exp \left( -k \psi (\scan (\br ) /\sqrt{k}, \lambda ) \right)~, \quad \forall \lambda >0\ .
\]
Note that $x\leq \sinh (x) \leq e^x /2$ and $|\bbS | \leq N$, hence
\begin{equation}\label{Rank_eqn:small_bound}
\mathfrak{P}(\br ) \leq N \exp \left( -\lambda \sqrt{k}\, \scan (\br ) + \frac{\lambda k N}{2} - k\log (\lambda N) \right)~, \quad \forall \lambda >0\ .
\end{equation}
The next step is to remove the conditioning $\bR =\br$ and bound $\scan (\bR )$. Recall $\scan(\bR ) \geq Y_\cS (\bR ) - \sqrt{k} \frac{N+1}{2}$, where $\cS$ is the anomalous interval. 
% Assume for now that $\theta_i = \theta_\ddag$ for all $i\in \cS$. 
As in the proof of Theorem~\ref{thm:rank_test} we use Lemma~\ref{lem:rank-sum} to evaluate the terms $\E (Y_\cS (\bR ))$ and $\Var (Y_\cS (\bR ))$. We have
\[
\E ( Y_\cS (\bR )) = \sqrt{k} (N-k) ( \bar{p}_\cS - 1/2 )) + \sqrt{k} \frac{N+1}{2} \ ,
\]
where we use the shorthand notation $\bar{p}_\cS = \tfrac{1}{|\cS |} \sum_{v\in \cS} p_v$.  For the variance term, recalling the definition of $\lambda_v$ from Lemma~\ref{lem:rank-sum}, we note that $\lambda_v \leq p_v$. Hence
\[
\Var (R_v) = (\lambda_v - p_v^2 ) n^2 +O(kN) \leq p_v (1-p_v ) N^2 +O(kN) \leq (1-p_v ) N^2 +O(kN) \ .
\]
Also using $\Cov (R_v ,R_w) = O(N)$, we get
\[
\Var (Y_\cS (\bR )) \leq (1- \bar{p}_\cS ) N^2 + O(kN) \ .
\]

According to our assumption, there exists a sequence $\omega_N \to \infty$ such that
\[
\bar{p}_\cS \geq 1-\omega_N^{-1} N^{-2/k} \ .
\]
For reasons that become apparent at the end of the proof, we choose $\omega_N \to \infty$ not too fast (for instance $\omega_N \leq \log N$ suffices). Using Chebyshev's inequality we get
\begin{align*}
\P & \Bigg( Y_\cS (\bR ) - \sqrt{k} \tfrac{N+1}{2} \leq \sqrt{k} (N-k) \left( \tfrac{1}{2}- \omega_N^{-1/4} N^{-1/k} \right) \Bigg) \\
& = \P \left( Y_\cS (\bR ) - \E (Y_\cS (\bR )) \leq \sqrt{k} (N-k) \left( 1- \omega_N^{-1/4} N^{-1/k} -\bar{p}_\cS \right) \right) \\
& \leq \P \left( Y_\cS (\bR ) - \E (Y_\cS (\bR )) \leq - \sqrt{k} (N-k) \left( \omega_N^{-1/4} N^{-1/k} - \omega_N^{-1} N^{-2/k} \right) \right) \\
& \leq \P \left( \left| Y_\cS (\bR ) - \E (Y_\cS (\bR )) \right| \geq \sqrt{k} (N-k) \left( \omega_N^{-1/4} N^{-1/k} - \omega_N^{-1} N^{-2/k} \right) \right) \\
& \leq \frac{N^2 \omega_N^{-1} N^{-2/k} + O(kN)}{k(N-k)^2 \left( \omega_N^{-1/4} N^{-1/k} - \omega_N^{-1} N^{-2/k} \right)^2} \leq \frac{4 N^2 \omega_N^{-1} N^{-2/k} + O(kN)}{k(N-k)^2 \omega_N^{-1/2} N^{-2/k}} \to 0 \ ,
\end{align*}
where the last inequality follows because $\omega_N^{-1/4} N^{-1/k} - \omega_N^{-1} N^{-2/k} \geq \omega_N^{-1/4} N^{-1/k}/2$ eventually as $N\to \infty$. Hence,
\[
\scan (\bR ) \geq \sqrt{k} (N-k) \left( \tfrac{1}{2}- \omega_N^{-1/4} N^{-1/k} \right) \ ,
\]
with probability converging to 1 as $N\to \infty$. Using this with \eqref{Rank_eqn:small_bound} we get
\[
\log \mathfrak{P}(\bR ) \leq \log N + \frac{\lambda k^2}{2} + \lambda k (N-k) \omega_N^{-1/4} N^{-1/k} - k\log (\lambda N)~, \quad \forall \lambda >0\ ,
\]
with probability tending to 1. Choosing $\lambda = \omega_N^{1/4} N^{1/k}/N$ we get
\[
\log \mathfrak{P}(\bR ) \leq \frac{\omega_N^{1/4} N^{1/k}}{N} k^2 + \frac{N-k}{N} k - \frac{k}{4} \log \omega_N \to -\infty \ ,
\]
with probability going to 1, where we used that $\omega_N$ grows slowly enough for the first term to vanish.

\medskip\noindent {\em Condition (ii).} 
We can mimic the arguments above. Suppose $k = c \log N$ with arbitrary $c>0$ and $\bar{p}_\cS = 1-(1-\delta)\exp (-\tfrac{c+1}{c})):= 1- (1-\delta )f(c)$ with some $\delta>0$. As before, using Chebyshev's inequality we can show that 
\[
\scan (\bR ) \geq \sqrt{k} (N-k) \left( \frac{1}{2} - \left( 1-\frac{\delta}{2} \right) f(c) \right) \ ,
\] 
with probability tending to 1 as $N\to \infty$. Plugging this into \eqref{Rank_eqn:small_bound}, choosing $\lambda = 1/(N f(c))$ we get
\[
\log \mathfrak{P}(\bR ) \leq \log N + \frac{k^2}{2 f(c) N} + \frac{k (N-k) (1-\tfrac{\delta}{2})}{N} - k \log f(c) \ ,
\]
with probability going to 1 as $N\to \infty$. Plugging in $k=c \log n$ and $f(c) = \exp (-\tfrac{c+1}{c})$ we see that the log of the $p$-value goes to $-\infty$, which is what we wanted to show.

\subsection{Additional results}
\subsubsection{Sketch proof of Lemma~\ref{lem:rank-sum}} \label{sec:proof-rank-sum}

First, assume that there are no ties in the ranks, with probability one. Note that we can write
\[
R_i = 1+\sum_{j\in [n],j\neq i} \IND{Z_i > Z_j} = 1+\sum_{j\in[s],j\neq i} \IND{Z_i > Z_j} + \sum_{j\notin [s],j\neq i} \IND{Z_i > Z_j} \ .
\]
Taking expectation yields
\[
\E (R_i) = \left\{
\begin{array}{ll}
1+(n-s) p_{i,0} + \displaystyle{\sum_{j\in [s],j \neq i}} p_{i,j} & \textrm{, when $i\in [s]$,}\\
\frac{n+s+1}{2} - \displaystyle{\sum_{j\in [s]}} p_{j,0} & \textrm{, when $i\notin [s]$.}
\end{array} \right.
\]
since $\P (Z_i = Z_j)=0$ for $i\neq j$ when there are no ties. The variance and covariance terms can be worked out using the same representation of the ranks as above, but we omit these straightforward computations for the sake of space.

In case of ties, to keep the presentation simple, assume that the distributions of $\{ Z_i \}_{i\in [n]}$ are supported on $\mathbb{Z}$. Then randomly breaking ties in the ranks amounts to using the following procedure. Let $\{ \epsilon_i \}_{i\in [n]}$ be independent and uniformly distributed on $(-c,c)$ with $c\leq 1/2$, also independent from $\{ Z_i \}_{i\in [n]}$. Consider $Z'_i = Z_i + \epsilon_i, \ i\in [n]$ and let $R'_i$ be the rank of $X'_i$ in the combined sample $\{ Z'_i \}_{i\in [n]}$. Then the joint distribution of $\{ R'_i \}_{i\in [n]}$ is the same as that of $\{ R_i \}_{i\in [m]}$ when ties are broken randomly.

For instance, for $i\notin [s]$
\begin{align*}
\E (R'_i) & = \tfrac{n+s+1}{2} - \sum_{j\in [s]} \P (Z'_i > Z'_j) \\
& = \tfrac{n+s+1}{2} - \sum_{j\in [s]} \left( \P (Z_i > Z_j) + \P (\epsilon_i > \epsilon_j |Z_i=Z_j) \P (Z_i = Z_j) \right) \\
& = \tfrac{n+s+1}{2} - \sum_{j\in [s]} p_{j,0} \ .
\end{align*}
The rest of the claims can be worked out similarly.

Finally, when $Z_i$ have arbitrary distributions a similar method can be applied, although it requires a bit more care and one needs to take $c$ approaching zero.

\subsubsection{Derivation of $\Upsilon_0$ in the normal location model} \label{sec:C0}

Assume the normal model where $F_\theta = \cN(\theta, 1)$.  
For this case we can simply compute $\Upsilon_0$.
Since there are no ties with probability 1, we have
\[
\Upsilon_0 = \E (X \IND{X>Y} ) = \int_{- \infty}^\infty \int_x^\infty u f_0(u) {\rm d} u f_0 (x) {\rm d} x .
\]
Considering the inner integral we have
\[
\int_x^\infty u f_0(u) {\rm d}  u = \frac{1}{\sqrt{2\pi}} \int_x^\infty u e^{-u^2 /2} {\rm d} u = \frac{1}{\sqrt{2\pi}} e^{-x^2 /2} = f_0 (x) \ .
\]
Hence
\[
\Upsilon_0 = \int_{-\infty}^\infty f_0 (x) = \int_{-\infty}^\infty \frac{1}{2\pi} e^{-x^2} {\rm d} x = \frac{1}{2\sqrt{\pi}} \ .
\]
Therefore we conclude that $1/(2\sqrt{3}\Upsilon_0)=\sqrt{\pi/3}$.

\subsection*{Acknowledgments}

We would like to thank Daniel Neill for references on fast scans; Martin Kulldorff for references on permutation scans, for bringing the reference \citep{jung2015nonparametric} to our attention, and for helpful comments; Bing Zhou for help with the real data application of \secref{CNV}; and Tengyao Wang and Richard Samworth for discussions about the properties of $\Upsilon_0$.  
Comments of an anonymous referee helped improve the narrative and presentation.  
The basic concept of this paper arose out of conversations between some of the authors at the 2013 Mathematical Statistics conference in Luminy, France. 
This work was partially supported by a grant from the US National Science Foundation (DMS 1223137) and a grant from the {\em Nederlandse organisatie voor Wetenschappelijk Onderzoek} (NWO 613.001.114).

\bibliographystyle{chicago}
\bibliography{rank-scan}

\end{document}